\documentclass[jcp,twocolumn,amsmath,amssymb,letterpaper,superscriptaddress,floatfix]{revtex4-1}

\usepackage{graphicx}% Include figure files
\usepackage{dcolumn}% Align table columns on decimal point
\usepackage{bm}% bold math
\usepackage{amsmath,amssymb,amsfonts,units,xspace}
\usepackage{color}
\usepackage[english]{babel} 
\graphicspath{{figures/}}

\usepackage{localdefs}

\newcommand{\eg}{{e.g., }}

\newcommand{\kT}{k_{\rm B}T}
\newcommand{\tm}{\textcolor{magenta}}
\newcommand{\tb}{\textcolor{blue}}

\begin{document}

%\preprint{AIP/123-QED}

\title[Structure and dynamics of a layer 
of sedimented Brownian particles]{Structure and dynamics of a layer 
of sedimented particles}
%\thanks{Footnote to title of article.}

\author{Adar Sonn--Segev}
\affiliation{Raymond and Beverly Sackler School of Chemistry, Tel Aviv University, Tel Aviv 6997801, Israel}

\author{Jerzy B\l{}awzdziewicz} 
\affiliation{Department of Mechanical Engineering, Texas Tech University, 7th and Boston, Lubbock, Texas 79409, USA}
\author{Eligiusz Wajnryb} 
\affiliation{ Institute of Fundamental Technological Research, Polish Academy of Sciences, Pawi\'nskiego 5B, Warsaw 02-106, Poland}
\author{Maria L. Ekiel--Je\.{z}ewska} 
\affiliation{ Institute of Fundamental Technological Research, Polish Academy of Sciences, Pawi\'nskiego 5B, Warsaw 02-106, Poland}
\author{Haim Diamant} 
\affiliation{Raymond and Beverly Sackler School of Chemistry, Tel Aviv
  University, Tel Aviv 6997801, Israel}

\author{Yael Roichman} 
\affiliation{Raymond and Beverly Sackler School of Chemistry, Tel Aviv
  University, Tel Aviv 6997801, Israel} \email{roichman@tau.ac.il}

\date{\today}% It is always \today, today,
             %  but any date may be explicitly specified

\begin{abstract}

We investigate experimentally and theoretically thin layers of colloid
particles held adjacent to a solid substrate by
gravity. Epifluorescence, confocal, and holographic microscopy,
combined with Monte Carlo and hydrodynamic simulations, are applied to
infer the height distribution function of particles above the surface,
and their diffusion coefficient parallel to it. As the particle area
fraction is increased, the height distribution becomes bimodal,
indicating the formation of a distinct second layer.  In our theory we
treat the suspension as a series of weakly coupled
quasi-two-dimensional layers in equilibrium with respect to particle
exchange.  We experimentally, numerically, and theoretically study the
changing occupancies of the layers as the area fraction is
increased. The decrease of the particle diffusion coefficient with
concentration is found to be weakened by the layering. We demonstrate
that particle polydispersity strongly affects the properties of the
sedimented layer, because of particle size segregation due to gravity.

\end{abstract}

\pacs{Valid PACS appear here}% PACS, the Physics and Astronomy
                             % Classification Scheme.

%Use showkeys class option if keyword
                              %display desired
\maketitle

\section{Introduction}
\label{sec:intro}

Being relevant to a wide range of practical scenarios, the behavior of
colloid suspensions near solid surfaces has been thoroughly studied
over the years. This research effort consists of several bodies of
work, for each of which we can give only a few representative
references. The first category of papers concerns the disruption of the
structural isotropy of a three-dimensional (3D) fluid suspension by
the surface, \eg the formation of a layered structure decaying away
from the surface under equilibrium
\cite{VanWinkle1988,GonzalezMozuelos1991} and nonequilibrium
\cite{Zurita_Gotor-Blawzdziewicz-Wajnryb:2012} conditions. Another
category addresses the effect of the anisotropic geometry on particle
dynamics near a single planar surface\,---\,for isolated particles
\cite{Happel,Perkins1992,%
Cichocki-Jones:1998,%
Walz-Suresh:1995,%
Prieve1999,%
Prieve2000,CarbajalTinoco2007,%
Blawzdziewicz2010%
},
particle pairs 
\cite{Happel,Dufresne2000,Cichocki2007,%
Zurita_Gotor-Blawzdziewicz-Wajnryb:2007b}, 
and a 3D suspension adjacent to a surface
\cite{Anekal2006,Michailidou2009,%
Cichocki-Wajnryb-Blawzdziewicz-Dhont-Lang:2010,%
Dhont2012,%
Michailidou2013}.  

Regarding
quasi-two-dimensional (quasi-2D) layers of particles, most studies have
considered the confinement of suspensions between two rigid
surfaces. This research addressed structural properties of such
confined suspensions
\cite{CarbajalTinoco1996,Schmidt1997,Zangi2000,Frydel2003,Han2008},
and the dynamics of single particles
\cite{Lin2000,Dufresne2001,Ekiel_Jezewska-Wajnryb-Blawzdziewicz-Feuillebois:2008},
particle pairs \cite{Cui2004,Bhattacharya2005b}, and concentrated quasi-2D suspensions
\cite{Diamant2005A,%
Blawzdziewicz-Wajnryb:2012,%
Baron-Blawzdziewicz-Wajnryb:2008}. 
Another type of quasi-2D suspensions has also been studied, where a
particle layer is confined to a fluid interface
\cite{Lin1995,Cichocki2004,Peng2009,Zhang2014}.

In cases where the surface attracts the particles and the suspension
is sufficiently dilute, the system can contain a single layer of
surface-associated particles in contact with a practically
particle-free solvent \cite{GonzalezMozuelos1991}. A single layer can
also form as a result of gravitational settling of particles toward a
horizontal wall.  This scenario is studied in the present work.

Sedimented colloidal particles undergo random Brownian displacements,
which results in diffusive broadening of the fluctuating particle
layer.  The width of the particle height distribution above the bottom
surface is characterized by the sedimentation length $l$, i.e., the
height at which the gravitational energy of a particle equals its
thermal energy.  The dynamics and height distribution of individual
sedimented particles above the bottom surface were studied in
Refs.\ \citenum{Walz-Suresh:1995,Prieve1999,Prieve2000} using total
internal reflection microscopy.  Particle monolayers at higher
densities were investigated experimentally for a system in which the
sedimentation length is much smaller than the particle
diameter \cite{Skinner2010}.  It was shown that at high area fractions
the suspension can assemble into quasi-2D colloidal crystals, but
formation of a nonuniform vertical microstructure was not observed,
because of the small sedimentation length.

Here we are interested in the structure and dynamics of a
surface-associated layer for which the sedimentation length is
comparable to the particle diameter.  We focus on the effects of the
suspension concentration on the statistical height distribution of
particles and their diffusion coefficient. Unlike the quasi-2D
suspensions confined between two surfaces or adsorbed at a fluid
interface (which restricts particle configurations and motions in two
directions) in the present system no constraints are imposed on the
distance between the particles and the single wall.  Thus, at
sufficiently high area fractions, particles form a nontrivial
stratified microstructure.  This microstructure and its effect on
particle dynamics are analyzed in our paper.

The article is organized as follows. Section~\ref{sec:exp_met}
describes the experimental methods used to prepare the system, image
the particles, and analyze the extracted data. In
Sec.\ \ref{sec:num_met} we describe the theoretical background and
numerical methods used to perform the simulations.  In
Sec.\ \ref{sec:structure} we present the results concerning the
equilibrium structure of the quasi-2D suspension observed in planes
parallel to the bottom surface (the quasi-2D radial distribution) and
in the direction perpendicular to it (the height distribution).
Section~\ref{sec:dynamics} addresses the diffusion of particles
parallel to the surface, as affected by the surface proximity.  We
discuss our findings in Sec.\ \ref{sec:discuss}.

\section{Experimental methods}
\label{sec:exp_met}

\subsection{Quasi-2D system of sedimented Brownian spheres}
\label{Quasi-2D system of sedimented Brownian spheres}

\begin{figure}
\centering
\includegraphics[clip=true,trim=0 0 0 0,scale=0.4]{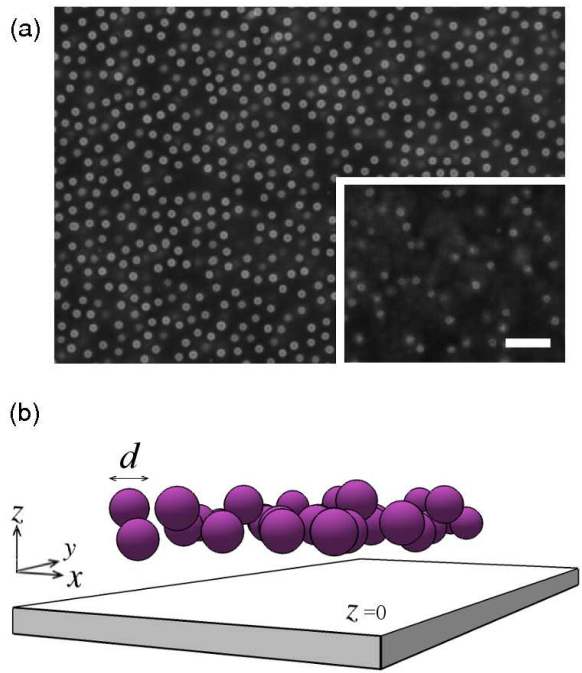}
\caption{(a) Images of fluorescent $1.5~\mu$m-diameter silica spheres
  suspended in water, taken after the particles sedimented to create a
  quasi-2D suspension at area fraction $\phi=0.49$. Large (small)
  image corresponds to a typical image of the first (second)
  layer. Scale bar = $5~\mu$m. (b) Schematic view of the system and
  its parameters. }
\label{fig:setup}
\end{figure}

Quasi-2D colloidal layers are created by placing a suspension of
colloidal silica spheres in a glass sample cell $\sim 150\,\mu$m
high. The particles are then allowed to sediment and equilibrate for
30 minutes at a temperature of approximately 24$^\text{o}$C before
measurements start (Fig.~\ref{fig:setup}). We use green fluorescent
monodisperse, negatively charged silica particles (Kisker Biotech,
PSI-G1.5 Lot \#GK0090642T) with diameter $d=1.50\pm0.15\,\mu$m, and
mass density $\massDensity=2.0$ g/cm$^3$. Monolayers of area
fraction $0<\phi\le0.62$ are prepared by diluting the original
suspension with double distilled water (DDW, 18~M$\Omega$), without
and with the addition of salt at a concentration $\KCl=0.01\Mol$. The
sample walls are cleaned and slightly charged by plasma etching to
avoid particle attachment to the bottom wall of the cell.  We observe
that the aqueous medium above the colloidal monolayer is free of
colloids. Since the particles are floating right above the bottom
wall, we can treat the upper wall as a distant boundary.

\subsection{Imaging techniques}

Particle position and motion in the $x$--$y$ plane, perpendicular to the
optical axis, are observed using epifluorescence microscopy (Olympus
IX71). Images are captured at a rate of $70$~fps by a CMOS camera
(Gazelle, Point Grey Research). We use in-line holographic microscopy
to image the dynamics of particles in three dimensions in dilute
samples \cite{Kapfenberger2013}. This imaging technique uses a collimated coherent light
source (DPSS, Coherent, $\lambda=532$nm) to illuminate a sample
mounted on a microscope. The light scattered from the sample
interferes with the light passing through it, to form a hologram in the
image plane. We reconstruct the light field passing through the sample
by Rayleigh-Sommerfeld back-propagation and extract from it the particle
location in three
dimensions \cite{Lee07a,Cheong2010b,Kapfenberger2013}. For holographic
imaging measurements we use non-fluorescent silica particles with the
same diameter ($d=1.50~\pm~0.08~\mu$m, Polysciences Inc.).
Additional details of the setup and measurement methods can be found
elsewhere \cite{Kapfenberger2013}.

\begin{figure}
\centering
\includegraphics[clip=true,scale=0.5]{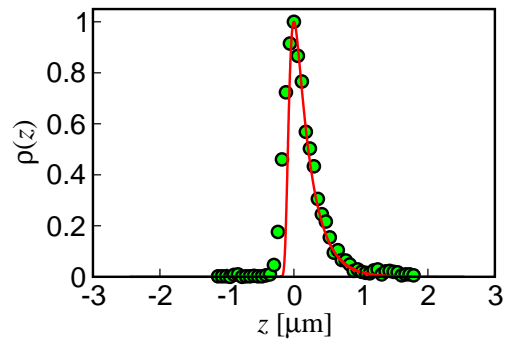}
\caption{Holographic imaging. Height probability distribution of a
  single sphere (in salt-free water) shifted to the maximum value and
  fitted to the Boltzmann probability distribution \eqref{low-density
    probability distribution} with the particle--wall potential
  (\ref{Eq:boltzmann}).}
\label{fig:boltzman}
\end{figure}

We use confocal imaging to monitor particle positions in a dense layer
in three dimensions. Our spinning disc confocal imaging system (Andor,
Revolution XD) includes a Yokogawa (CSU-X1) spinning disc, and an
Andor (iXon 897) EM-CCD camera. An objective lens (Olympus, x60,
NA=1.1, water immersion) mounted on a piezoelectric scanner (Physik
Instrumente, Pifoc P-721.LLQ) is used to scan the sample in the $z$
axis, with a step size of 100 nm.
 
\subsection{Height calibration}
\label{Height calibration}

A suspended tracer particle
is subject to electrostatic and gravitational forces in addition to
thermal fluctuations, affecting its height distribution
\cite{Prieve1999}. 
 The particle potential energy can be described as
 
\begin{equation}
U~= ~mgz+B\mathrm{e}^{-(z-d/2)/\lambda},
\label{Eq:boltzmann}
\end{equation}
where $z$ is the vertical position of the tracer, $g$ is the
gravitational acceleration,
\begin{equation}
\label{m}
m=\textstyle\frac{\pi}{6}\Delta\massDensity d^3
\end{equation}
is the buoyant mass of the tracer ($\Delta \massDensity$ is the mass
density difference between silica and water), $\lambda$ is the Debye
screening length, and the amplitude $B$ depends on $\lambda$ and the
surface charges of both particle and glass surfaces.  The
corresponding probability distribution of the particle height $z$ is
\begin{equation}
\label{low-density probability distribution}
\particleDistribution(z)=\particleDistributionNorm\mathrm{e}^{-U(z)/\kT},
\end{equation}
where $\kT$ is the thermal energy and $\particleDistributionNorm$ is
the normalization constant.

The height distribution of a single particle above the sample's bottom
was obtained from very dilute suspensions, using in-line holographic
imaging \cite{Lee07a,Cheong2010b,Kapfenberger2013} (see
Fig.~\ref{fig:boltzman}). Our holographic measurements provide values
of relative particle positions, but not the absolute particle heights
with respect to the bottom wall.  We thus set the peak position to
$z=0$ and focus on the height relative to this reference plane. The
exponential decay on the right side of the probability-density peak is
governed by a decay length,
\begin{equation}
l=\frac{k_BT}{mg}\label{ldef}
\end{equation}
(the sedimentation length), resulting from the competition between
gravity and thermal forces. The exponential-decay length determined
from the holographic measurements agrees well with the calculated
sedimentation length \eqref{ldef}, without any fitting parameters (see
Fig.~\ref{fig:boltzman}). 
   The electrostatic term of the probability-density, which controls the steep rise of the probability, affects mostly the peak position rather than its shape. Since we shifted the peak position to $z=0$, the fitting of the entire probability-density using Eqs.\ \eqref{Eq:boltzmann} and \eqref{low-density probability
      distribution} was insensitive to the value of $B$. Reasonable fits were obtained for $\lambda$ in the range of $\lambda \sim 40-70$ nm. Better estimations of $B$ and $\lambda$ are given in Sec.~\ref{Mean particle height at low area fractions}, using mobility measurements.
       
\begin{figure}
\centering \includegraphics[clip=true,scale=0.5]{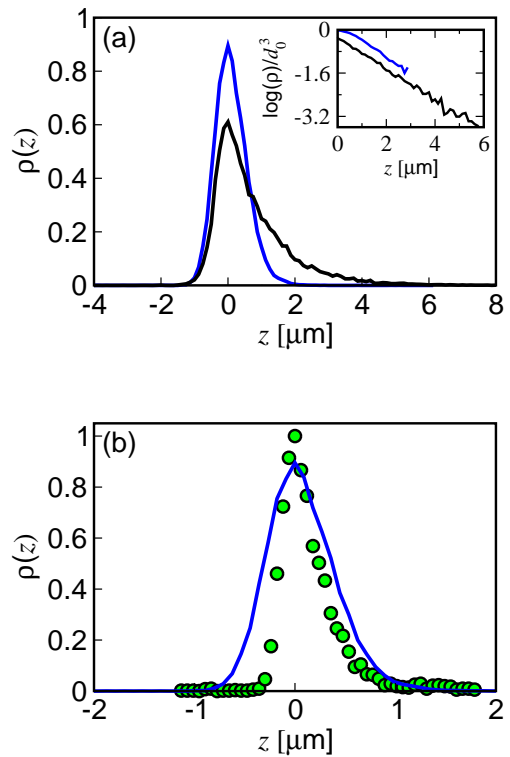}
\caption{(a) Confocal imaging. Height probability distribution of
  silica particles at $\phi<0.003$ with diameters of $d_0=1.5~\mu$m
  (blue) and $d_0=1.0~\mu$m (black). Inset: Logarithm of the
  probability distributions scaled by $d_0^3$ in units of
  $10^2~\mu$m$^{-3}$; as expected, the two curves have approximately
  the same slope, which is used to calibrate the confocal height
  measurements.  (b) Height probability distribution of silica
  particles with diameter $d_0=1.5~\mu$m extracted from holographic
  imaging (green circles, see Fig.~\ref{fig:boltzman}) and confocal
  imaging (blue line). The suspensions in both figures were with no added salt, $\KCl=0 \Mol$.  }
\label{fig:pvsz}
\end{figure}

The applicability of the holographic imaging is limited to low-density
suspensions, whereas the confocal imaging can be also used at higher
concentrations. On the other hand, confocal height measurements suffer
from spherical aberrations due to multiple changes in refractive index
in the imaging path. This leads to a systematic error in measuring
$z$, which can be eliminated by proper calibration. We calibrate the
confocal measurement of the relative vertical particle positions by
requiring the exponential decay of the height distribution to agree
with the known, and verified, value of $l$.

In Fig.~\ref{fig:pvsz}\subfig{a} we show the particle-height
distribution $\particleDistribution(z)$ at $\phi<0.003$ for two
different particle sizes ($d_0=1.0,1.5~\mu$m).  The distributions are
shifted so that the highest probability is located at $z=0$. Scaling
the logarithm of the distributions by $d_0^3$ [inset of
  Fig.~\ref{fig:pvsz}\subfig{a}] shows that the normalized decay
constants for the two particle sizes have approximately the same
value, from which we calibrate the confocal microscope's height
measurements.  In Fig.~\ref{fig:pvsz}\subfig{b}, the height
distributions extracted by the two methods (holographic and confocal
imaging) are overlaid. This figure emphasizes the higher accuracy of
holographic imaging over confocal imaging, especially around $z=0$,
where the increase in distribution should be very steep
\cite{Prieve1999, Kapfenberger2013}.  The difference between the
curves can also be attributed to polydispersity, since the holographic
imaging is a single-particle measurement while the confocal imaging is
a multiple-particle measurement, and its corresponding curve
represents an average over $\sim$40 particles.

\section{Numerical methods}
\label{sec:num_met}

\subsection{The system}
\label{The system - numerical}

\subsubsection{Particles and their interactions}
\label{Particles and their interactions}

Silica particles are modeled as Brownian hard spheres with or without
electrostatic repulsion (depending on the salt concentration), immersed in
a fluid of viscosity $\eta$.  The bottom wall is treated as an
infinite hard planar surface.  Creeping-flow conditions and no slip
boundary conditions at the particle surfaces and at the wall are
assumed.

In a salt solution with $\KCl=0.01\Mol$, the Debye length is only
about 5 nm, and therefore electrostatic interactions are screened out.
The particles thus interact only via infinite hard-core
particle--particle and particle--wall potentials and the gravity
potential $mgz$, and no other potential forces are involved.  The
strength of the gravity force is described by the sedimentation length
\eqref{ldef}.

In addition to the hard-core repulsion, in DDW with no added salt
($\KCl=0\Mol$) particles are assumed to also interact via particle--wall and
particle--particle Debye--H\"uckel potentials,
\begin{equation}
 V(z)=Be^{-(z-d/2)/\lambda},\label{V}
\end{equation}
and 
\begin{equation}
 V'(r)=B'e^{-(r-d)/\lambda},\label{V'}
\end{equation}
where $\lambda$ is the Debye screening length, $B$ and $B'$ are the
potential amplitudes, and $r$ is the distance between the particle
centers.  The consideration of Debye--H\"uckel potentials in the
salt-free case is based on our experimental measurement $\lambda \sim
60$ nm.  A finite Debye screening length in DDW stems from the
presence of residual ions in the solution \cite{Behrens2001}.

\subsubsection{Suspension polydispersity}
\label{Suspension polydispersity}
 
To determine the effects of the suspension polydispersity on the
near-wall microstructure and dynamics, we have performed numerical
simulations for a hard-sphere (HS)  system with a Gaussian
distribution of particle diameters,
\begin{equation}
\label{Particle size distribution}
p(d)=
\frac{1}{(2\pi\sigma^2)^{1/2}}\exp\left[-\frac{(d-\daver)^2}{2\sigma^2}\right],
\end{equation}
where $d$ and $\daver$ are the actual and average particle diameters,
and $\sigma$ is the standard deviation.  All the particles have the
same mass density $\massDensity$; hence, particles of different sizes
have different buoyant masses and different sedimentation lengths
\eqref{ldef}.  The dimensionless sedimentation length based on the
average particle diameter $d_0$, is defined as
\begin{equation}
\label{dimensionless sedimentation constant}
 \frac{l_0}{\daver}=\frac{k_BT}{\maver g\daver},
\end{equation}
where 
\begin{equation}
 \maver=\frac{\pi}{6} \daver^3\,\Delta\massDensity.
\end{equation}
The area fraction $\phi$ based on the average particle diameter
$\daver$ is
\begin{equation}
 \phi=\textstyle\frac{1}{4}\pi n\daver^2,\label{phi}
\end{equation}
where $n$ is the number of particles per unit area.  Since the
particles are free to move in the $z$ direction, the area fraction
$\phi$ can exceed 1.

\subsubsection{System parameters}

The simulations were carried out for the following system parameters:
For the dimensionless sedimentation length \eqref{dimensionless
  sedimentation constant} we use the value
\begin{equation}
\label{value of dimensionless sedimentation constant}
\frac{l_0}{\daver}=0.158,
\end{equation}
calculated from the particle size and density.  Based on the
comparison between the calculated and measured values of the
equilibrium average of the lateral self-diffusion coefficient for
isolated particles in DDW,  we estimate that
the Debye length and the amplitude of particle--wall electrostatic
repulsion are
\begin{equation}
\label{Debye parameters}
\lambda/d=0.03,\qquad \frac{B}{\kT}=10.
\end{equation}  
These values are used for salt-free suspensions at all suspension
concentrations.  Assuming that the charge densities of the particle and
wall surfaces are similar, we take
\begin{equation}
\label{Debye B particle}
B'=B/2,
\end{equation}
for the interparticle-potential amplitude, as follows from the Derjaguin approximation \cite{Israelachvili}.

The simulations were performed in the range of area fractions
$\phi\le1.2$.  For polydisperse HS systems the calculations
were carried out for $\sigma/d_0=0.10$, 0.15, 0.20, and 0.25 (we
estimate that $0.10 < \sigma/d_0 < 0.15$ for the silica particles used
in the experiments).  For particles interacting via the Debye--H\"uckel
potentials \eqref{V} and \eqref{V'} only monodisperse suspensions were
considered.

\subsection{Evaluation of the equilibrium distribution}
\label{Equilibrium distributions}

\subsubsection{Low density limit}
\label{Structure - low density limit}

For monodisperse suspensions at low particle concentrations, the
equilibrium particle distribution $\particleDistribution(z)$ is given
by the normalized Boltzmann factor \eqref{low-density probability
  distribution}.  To determine the particle distribution for a dilute
polydisperse suspension, the particle-size-dependent Boltzmann factor
for individual particles, $\particleDistribution_1(z;d)$, is
convoluted with the particle-size distribution \eqref{Particle size
  distribution}, 
\begin{equation}
\label{dilute polydisperse equilibrium distribution}
\particleDistribution(z)=\int_0^z \textrm{d}d\, 
p(d)\particleDistribution_1(z;d),
\end{equation}
For a HS system
\begin{equation}
\label{individual Boltzmann factor for hard spheres}
\particleDistribution_1(z;d)=l^{-1} \textrm{e}^{-(z-d/2)/l} \theta(z-d/2),
\end{equation}
according to equations \eqref{Eq:boltzmann}--\eqref{ldef}, where
$\theta(x)$ is the Heaviside step function, and the sedimentation
length $l$ is particle-size dependent due to the variation of particle
mass.

\subsubsection{Monte--Carlo simulations}
\label{Monte-Carlo simulations}

To determine the equilibrium microstructure of a sedimented suspension
at finite particle area fractions, equilibrium Monte--Carlo (MC)
simulations were performed for 2D-periodic arrays of spherical
particles in 3D space (with periodicity in the horizontal directions
$x$ and $y$ and the box size $L$).  The particles interact via
infinite hard-core repulsion and the pair-additive potential

\begin{equation}
U(\boldsymbol{X)}=\sum_{i=1}^{N}m_{i}gz_{i}+
\sum_{i=1}^{N}V(z_{i})+\frac{1}{2}\sum_{i=1}^{N}\sum_{j\ne i}^{N}V^{\prime
}(r_{ij}),
\end{equation}
which includes the gravity term and particle--wall and
particle--particle screened electrostatic potentials \eqref{V} and
\eqref{V'}.  Here $\boldsymbol{X}=(\mathbf{r}_{1},\ldots
,\mathbf{r}_{N})$ is the particle configuration (with $\mathbf{r}_i$
denoting the position of particle $i$), $z_i$ is the vertical
coordinate of particle $i$, and $r_{ij}=|\mathbf{r}_i-\mathbf{r}_j|$
is the relative particle distance.

A purely HS  system with $V=V'=0$ was modeled for monodisperse
particles and for polydisperse particles with the Gaussian size
distribution \eqref{Particle size distribution}.  For systems with
nonzero electrostatic repulsion only monodisperse particles were
considered.  

The initial configuration was prepared by placing $N=400$ particles
randomly in a vertical cuboid box with the square base $L$ and the
height $10L$.  The size $L$ of the 2D-periodic cell was determined to
obtain the required area fraction $\phi$ of the sedimented particle
layer.  The suspension was allowed to sediment by following the MC
random-walk dynamics in the configurational space $\boldsymbol{X}$
\cite{Frenkel-Smit:2002} (as described below).  After the equilibrium
state was reached, suspension properties were obtained by averaging
the quantities of interest over at least 200 independent
configurations.

Our adaptive simulation procedure was performed by repeating the MC
steps defined as follows:

\begin{itemize}
\item[\subfig{a}] A randomly selected particle $i$ is given a small
  random displacement, $\boldsymbol{r} _{i}\rightarrow
  \boldsymbol{r}_{i}^{\prime }=\boldsymbol{r}_{i}+\boldsymbol{ \Delta
  }$, where $\boldsymbol{\Delta }$ is chosen from a 3D Gaussian
  distribution with the standard deviation adaptively adjusted to the
  current mean gap between particles.  This displacement results in
  the change of the configuration from $\boldsymbol{X}$ to
  $\boldsymbol{X}^{\prime }$.

\item[\subfig{b}] According to the Metropolis detailed balance
  condition, the new configuration is accepted with the probability
\begin{equation}
\min \left( 1,\exp \left\{ -\left[ U(\boldsymbol{X}^{\prime })
-U(\boldsymbol{X})\right] /k_{B}T\right\} \right),
\end{equation}
provided that there is no particle--particle or particle--wall overlap.

\end{itemize}

To let the system reach an equilibrium state $\boldsymbol{X}_{1}$, the
MC step \subfig{a} and \subfig{b} is repeated $10^{5}N$ times.  The
next independent equilibrium configuration $\boldsymbol{X}_{n+1}$ is
obtained from the previous configuration $\boldsymbol{X}_{n}$ by
performing $10^{4}N$ MC steps.  The particle height distribution
$\particleDistribution(z)$ and other equilibrium quantities are
obtained by averaging over 200 independent configurations
$\boldsymbol{X}_{i}$.

\subsection{Hydrodynamics and self-diffusion}
\label{Hydrodynamics and self-diffusion}

\subsubsection{Low density limit}
In the absence of a wall, the self-diffusion coefficient of an
isolated solid sphere with diameter $d_0$ is given by the
Stokes--Einstein expression
\begin{equation}
 D_0=\frac{\kT}{3\pi \eta d_0}.
\end{equation}
The self-diffusion coefficient $D(z)$ of a sphere with diameter $d$ at a
distance $z$ from the wall is smaller by a factor 
\begin{equation}
\label{self diffusivity for single sphere in wall presence}
 \frac{D(z)}{D_0}=\frac{d_0}{d}\,\mu_{\parallel}(z/d),
\end{equation}
where the normalized mobility coefficient $\mu_{\parallel}$ depends on
the dimensionless particle position $z/d$ and no other parameters.
Relation \eqref{self diffusivity for single sphere in wall presence}
refers to the lateral component of the self-diffusion coefficient
(parallel to the wall), which was measured in our experiments.
However, an analogous expression also holds for the normal component.

For monodisperse particles in the dilute-suspension limit, the
effective self-diffusion coefficient $\Dself$ averaged across the
suspension layer is obtained by integrating \eqref{self diffusivity
  for single sphere in wall presence} with the Boltzmann distribution
\eqref{low-density probability distribution},
\begin{equation}
\label{monodisperse effective self-diffusivity}
\frac{\Dself}{D_0}=\int_{d/2}^\infty\textrm{d}z\, \particleDistribution(z)\mu_{\parallel}(z/d).
\end{equation}
For a polydisperse suspension, an additional average over the
particle-size distribution \eqref{Particle size distribution} is
needed, 
\begin{equation}
\label{polydisperse effective self-diffusivity}
\frac{\Dself}{D_0}=
  \int_0^\infty\textrm{d}d\,p(d)
   \int_{d/2}^\infty\textrm{d}z\,\particleDistribution_1(z;d)\mu_{\parallel}(z/d).
\end{equation}

The mobility coefficient $\mu_{\parallel}(z/d)$ was evaluated with
high accuracy using the \textsc{Hydromultipole} algorithm for a
particle near a single wall \cite{Cichocki-Jones:1998}.  The
integrals in Eqs.\ \eqref{monodisperse effective self-diffusivity}
and \eqref{polydisperse effective self-diffusivity} were performed
numerically using the Gauss method, with $\mu_{\parallel}(z/d)$
calculated by a series expansion.

\subsubsection{Computations for larger densities}

The effective self-diffusion coefficient for suspensions at higher
concentrations was evaluated using a periodic
version~\cite{Blawzdziewicz-Wajnryb:2008} of the
Cartesian-representation algorithm for a system of particles in a
parallel-wall channel
\cite{Bhattacharya-Blawzdziewicz-Wajnryb:2005a,Bhattacharya-Blawzdziewicz-Wajnryb:2005}.
In our approach, periodic boundary conditions in the lateral directions
are incorporated by splitting the flow reflected by the particles into
a short-range near-field contribution and a long-range asymptotic
Hele--Shaw component. The near-field contribution is summed explicitly
over neighboring periodic cells, and the Hele--Shaw component is
evaluated using Ewald summation method for a 2D harmonic
potential \cite{Cichocki-Felderhof:1989,Blawzdziewicz-Wajnryb:2008}.

The one-wall results were derived from the two-wall calculations using
an asymptotic procedure based on the observation that in the
particle-free part of the channel the velocity field tends to a
combination of a plug flow and a shear flow.  All other flow components
decay exponentially with the distance from the particle layer. The
one-wall results are obtained by eliminating the shear flow and
retaining only the plug flow generated by hydrodynamic forces induced
on the particles
\cite{Sadlej-Wajnryb-Blawzdziewicz-Ekiel_Jezewska-Adamczyk:2009}.  The
calculations were performed for the distance to the upper virtual wall
$H=10d_0$, which is sufficient to obtain highly accurate one-wall
results.

The self-diffusion coefficient is determined by averaging the trace of
the lateral translational--translational $N$-particle mobility,
evaluated using {\sc Hydromultipole} codes based on the above
algorithm, with the multipole truncation order $L=2$
\cite{Abade:2010}. The averaging was performed over equilibrium
configurations of $N=400$ particles in a 2D-periodically replicated
simulation cell.  Independent equilibrium configurations were
constructed using the MC technique described in Sec.\ \ref{Monte-Carlo
  simulations}.
\section{Structure of the quasi-2D suspensions}
\label{sec:structure}

\subsection{Experimental results}
\label{Structure - experimental results}

A typical image of our quasi-2D colloidal suspension is
shown in Fig.~\ref{fig:setup}.  For each area fraction $\phi$ and salt
concentration, the suspension can be characterized by the structure in
the $x$--$y$ plane (parallel to the cell floor) and the density profile in
the $z$ direction (perpendicular to the floor).  In this section we
discuss results of our measurements of the microstructure of a
sedimented particle layer.

\subsubsection{Mean particle height at low area fractions}
\label{Mean particle height at low area fractions}

As mentioned in Sec.\ \ref{Height calibration}, our imaging techniques
do not yield absolute particle heights. To estimate the mean particle
distance $z$ from the bottom wall (the mean height) in a dilute
suspension layer, we observe particle dynamics in the horizontal
directions, and compare measurement results with theoretical
calculations of the effect of the wall on the lateral particle
diffusion.  Using fluorescence imaging, we determine the projection of
particle trajectories onto the $x$--$y$ plane,
$\mathbf{r}_\parallel(t)$, and extract the effective self-diffusion
coefficient,
\begin{equation}
\label{self diffusivity from mean-square displacement}
\Dself=\langle \Delta \textbf{r}_\parallel^2
(\tau)\rangle/(4 \tau),
\end{equation}
where $\tau$ is the time interval.  The position-dependent diffusivity
$D(z)$ in the $x$--$y$ plane of a single particle near a planar wall
is given by the following expansion in the particle--wall distance \cite{Happel,Perkins1992},
\begin{eqnarray}
\frac{D(z)}{D_0} =   1 &-& \frac{9}{32}\frac{d}{z}
  +\frac{1}{64}\left( \frac{d}{z}\right)^3 \label{Eq:ds_zero}
  \\ \nonumber &-& \frac{45}{4096}\left( \frac{d}{z}\right)^4
  -\frac{1}{512}\left( \frac{d}{z}\right)^5,
\end{eqnarray}
where $z=0$ is the wall position. The expansion \eqref{Eq:ds_zero} is accurate within 5\% to 1\% as $z$ increases from $0.51d$ up to $d/2+2l$, in the range where sedimented particles spend most of
the time in a low-density suspension under equilibrium conditions.
Here $l\approx0.16d$ is the sedimentation length \eqref{ldef}.

From expression \eqref{Eq:ds_zero} and $\Dself$ extracted according to
\eqref{self diffusivity from mean-square displacement}, we can
calculate the suspension's mean distance from the wall (where $z$ in
\eqref{Eq:ds_zero} is replaced by a mean value $\langle z
\rangle$). This calculation holds in the limit $\phi \to
0$, where there are no particle-particle interactions. We measure
$\Dself$ from the particle trajectories, $\mathbf{r}_\parallel(t)$, in
extremely low area fraction solution, $\phi<0.003$ (in salt-free
water), and obtain a mean distance from the wall $\langle z \rangle=
1.1 \pm 0.1\,\mu$m, corresponding to a mean gap $\epsilon=z-d/2$ of
0.3--0.4\,$\mu$m between the particle surface and the wall.  We also
extract $\langle z\rangle$ for different salt concentrations by
extrapolating $\Dself$ (measured at various area fractions) to
$\phi=0$ (see Sec.~\ref{Dynamics - experimental results}), obtaining
$\langle z\rangle =0.95 \pm 0.05\,\mu$m for $\KCl=0.01\Mol$ and
$\langle z\rangle =1.11 \pm 0.05\,\mu$m for $\KCl=0\Mol$. The latter
matches the average height extracted from the diffusion of tracers in
the extremely low density suspension.

For $\lambda=5$ nm (added salt), the mean height calculated from the
Boltzmann distribution \eqref{low-density probability distribution} is
dominated by the exponential decay due to gravity and is practically
independent of $B$ in the particle--wall potential
\eqref{Eq:boltzmann}.  For $B=0$, using the particle mass as
determined from Eq.\ \eqref{m} with no fitting parameters, we get
$\langle z\rangle = 0.99$ $\mu$m, in agreement with the
diffusivity-based measurements of $\langle z\rangle$.  This result
confirms that in the added-salt case we can neglect the electrostatic
repulsion from the wall. For the salt-free case, taking $\lambda=50$
nm, we obtain $\langle z\rangle = 1.11$ $\mu$m for $B$ in the range
5--15 $\kT$. These values are consistent with those obtained by
fitting the measured height distribution to the theoretical expression
\eqref{Eq:boltzmann} (see Sec.~\ref{Height calibration}).

Since Eq.\ \eqref{Eq:ds_zero} does not include lubrication correction
for small particle--wall gaps $\epsilon$, it overpredicts $D(z)$
for $z<0.51d$; however, the accuracy of the approximation is sufficient
for the purpose of the present estimates.  In our calculations
discussed in Secs.\ \ref{sec:num_met} and \ref{Dynamics - numerical
  simulations}, highly accurate \textsc{Hydromultipole} results were
used instead of the far-field approximation \eqref{Eq:ds_zero}.

\begin{figure}
\centering
\includegraphics[clip=true,scale=0.5]{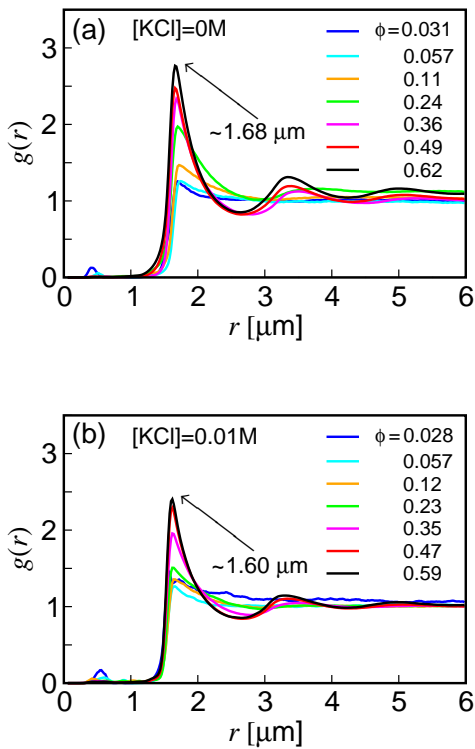}
\caption{Radial distribution function $g(r)$ in the $x$--$y$ plane for
  experiments in (a) salt-free and (b) salt-added ($\KCl=0.01\Mol$)
  water. The distribution $g(r)$ was calculated separately in the
  first and second layers (see Sec.~\ref{Occupation of particle layers}
  for the layer definition) and combined with appropriate weights. }
\label{fig:gofr}
\end{figure}

\subsubsection{Radial distribution in the horizontal plane}

To verify that no crystalline or hexatic structures are formed at
higher values of the area fraction, we evaluate from the experiment
the radial distribution function $g(r)$ and the full 2D pair
distribution $g(r,\theta)$ in the $x$--$y$ plane, for both the
salt-free and salt-added suspensions.  No dependence on $\theta$ was
found.  The radial distribution $g(r)$ for several values of the area
fraction $\phi$ is shown in Fig.\ \ref{fig:gofr}\subfig{a} for the
salt-free system and in \ref{fig:gofr}\subfig{b} for the salt-added
system.

For monodisperse hard spheres the first peak of $g(r)$ should
correspond to the diameter of the sphere. Our measurements show that
the first peak is at $r = 1.68~\mu$m for suspensions without salt and
at $r = 1.60~\mu$m for suspensions with $\KCl=0.01\Mol$. The
difference between these two numbers implies that the effective shell
around the particles in the salt-free samples is around ~40-50~nm,
which provides an estimation for the screening length in DDW without
the addition of salt. This estimate of $\lambda$ is consistent with
the other two mentioned above.

\subsubsection{Vertical density profile}

\begin{figure}[t]
\centering
\includegraphics[clip=true,scale=0.5]{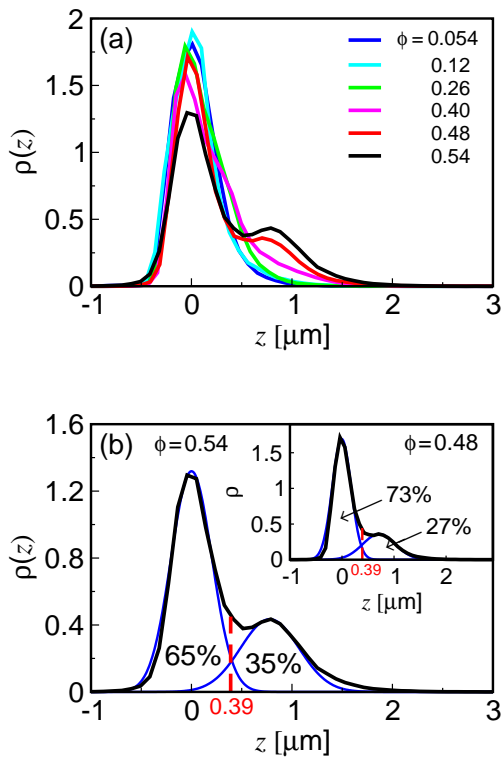}
\caption{(a) Height probability distribution of the silica colloids
  (in $\KCl=0.01\Mol$) for increasing area fraction reveals the
  formation of a second layer. Colors correspond to different area
  fractions (as labeled).  (b) For the most dense suspensions
  [$\phi=0.54$ and $\phi=0.48$ (inset)], the height distribution
  (black solid line) around the two peaks can be fitted to two
  Gaussian functions (blue lines). The intersection of the two
  Gaussians defines an effective boundary (red broken line) between
  the first and second layers; occupation percentages are indicated. }
\label{fig:zhist}
\end{figure}

The height distributions $\particleDistribution(z)$ of the silica
particles at different area fractions of the sedimented particle layer
were acquired using confocal imaging and conventional image analysis
\cite{Crocker1996}.  These distributions for salt-added suspensions
with $\KCl=0.01\Mol$ are plotted in Fig.~\ref{fig:zhist}\subfig{a} for
several values of the area-fraction $\phi$.  Since we cannot precisely
measure the position of the wall, the distributions are shifted so
that their first peak (close to the wall) is located at $z=0$.  These
distributions indicate the formation of a second layer of particles
for area fractions $\phi \gtrsim 0.26$.  The observed center of the
second layer is located $\Delta z\approx 0.75\,\mu$m above the center
of the first layer.  The layer separation is thus significantly
smaller than the expected separation $\Delta z\approx d=1.5\,\mu$m
(which is similar to the peak separation for the radial
distribution).  See further discussion in Secs.\ \ref{sec:discuss} and
Appendix \ref{appendix}.

\begin{figure}
\centering
\includegraphics[clip=true,scale=0.5]{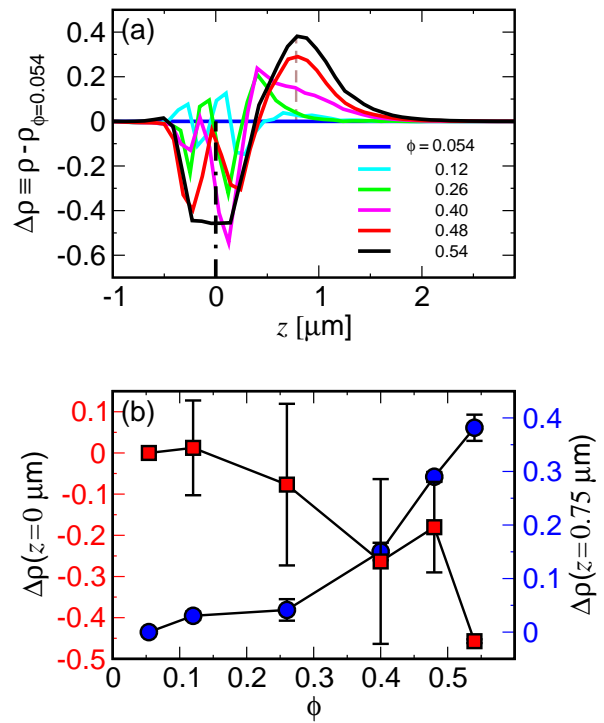}
\caption{(a) The difference between the height probability distribution at
  increasing area fractions and the distribution at the lowest area
  fraction $\phi =0.054$, $\Delta \rho \equiv \rho-\rho_{\,
    \phi=0.054}$ (in $\KCl=0.01\Mol$). Colors are as in
  Fig~\ref{fig:zhist}\subfig{a}. Gray (black) dashed line corresponds
  to $z=0.75~\mu$m ($z=0~\mu$m). (b) Values of $\Delta \rho$ at $z=0~\mu$m (red
  squares) and $z=0.75~\mu$m (blue circles) for all area
  fractions. Both plots exhibit a change in trend at area fraction
  $\phi \sim 0.3$.}
\label{fig:onset}
\end{figure}

To highlight the onset of the formation of the second layer, we look
at the subtraction of the height probability distribution of the
lowest area fraction from the distribution of all area fractions,
$\Delta \rho \equiv \rho-\rho_{\,\phi=0.054}$
[Fig.~\ref{fig:onset}\subfig{a}]. Two phenomena are expected when a
second layer is formed: (i) negative values at $z=0~\mu$m,
corresponding to a reduction in the fraction of particles populating
the first layer, (ii) positive and increasing values at $z=0.75~\mu$m,
corresponding to the formation and increasing population of the second
layer. The values of $\Delta\rho$ at $z=0$ and $0.75~\mu$m are plotted
in Fig.~\ref{fig:onset}\subfig{b}. The two expected phenomena are
observed at approximately $\phi \sim 0.3$, indicating the area
fraction above which a second layer becomes occupied. At area
fractions smaller than 0.3 we still obtain negative values of $\Delta
\rho$ at $z=0~\mu$m, and positive values at $z=0.75~\mu$m, however
these values are relatively low and can correspond to the broadening
of the exponential distribution due to increase in $\phi$.

\subsubsection{Particle-layer occupation fractions}
\label{Occupation of particle layers}

For the area fractions at which a clear second peak in the particle
distribution $\rho$ is seen in Fig.\ \ref{fig:zhist}\subfig{a} 
(i.e., for $\phi=0.48$ and $0.54$), we fit the area around
each peak to a Gaussian function and define the point of intersection
between the two Gaussians as the effective boundary between the two
layers. Figure~\ref{fig:zhist}\subfig{b} shows the two distributions
with the Gaussian fits and our definition of that boundary, which
turns out to be at a distance of $0.39 \pm 0.04~\mu$m above the peak
of the first layer in both densities. 

\begin{figure}[t]
\centering
\includegraphics[clip=true,scale=0.5]{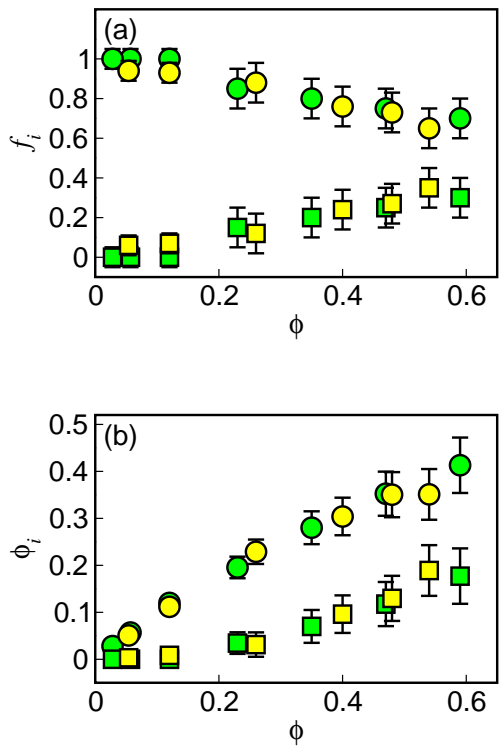}
\caption{Occupation fractions $f_i$ and area fractions $\phi_i$ of
  suspension layers as a function of the total area fraction $\phi$;
  circles (squares) correspond to the first (second) layer. (a) Occupation
  fractions for experiments in $\KCl=0.01\Mol$ extracted from the
  confocal height distribution curves (yellow) and from the 2D images
  (green), showing good agreement between the two methods.  (b) The
  same data replotted for the area fractions $\phi_1$ and $\phi_2$
  of the first and second layers. }
\label{fig:occ}
\end{figure}

Using this boundary, we evaluated the occupation fractions
$\occupationFraction{i}=\phi_i/\phi$ of the bottom ($i=1$) and top
layer ($i=2$), where $\phi_i$ is the area fraction of particles in
layer~$i$.  The results are shown in Fig.~\ref{fig:occ}\subfig{a} for
a suspension in $\KCl=0.01\Mol$ solution as a function of the total
area fraction $\phi$. As expected, the fraction of particles
populating the second layer grows as the total area fraction of
the suspension is increased.  

An additional independent measurement of the layer occupation
fractions is done using epifluorescence microscopy, which enables us
to image the different layers separately [see
  Fig.~\ref{fig:setup}\subfig{a}]. The occupation fraction of each
layer is determined by counting the number of particles observed
therein.  The occupation fractions measured using the epifluorescence
imaging technique are plotted in Fig.\ \ref{fig:occ}\subfig{a} along
with the results obtained from the confocal microscopy.  The two
methods yield similar results.

Alternatively, we can represent the layer-occupation results in terms
of the area fractions $\phi_1$ and $\phi_2$ of the first and second
layers [see Fig.~\ref{fig:occ}\subfig{b}].  Both $\phi_1$ and $\phi_2$
increase as $\phi$ is increased, and $\phi_1$ seems to
saturate at $\phi>0.45$.

\subsection{Numerical simulations}
\label{Structure - numerical simulations}

\begin{figure}
\includegraphics[width=\ffraction\textwidth]{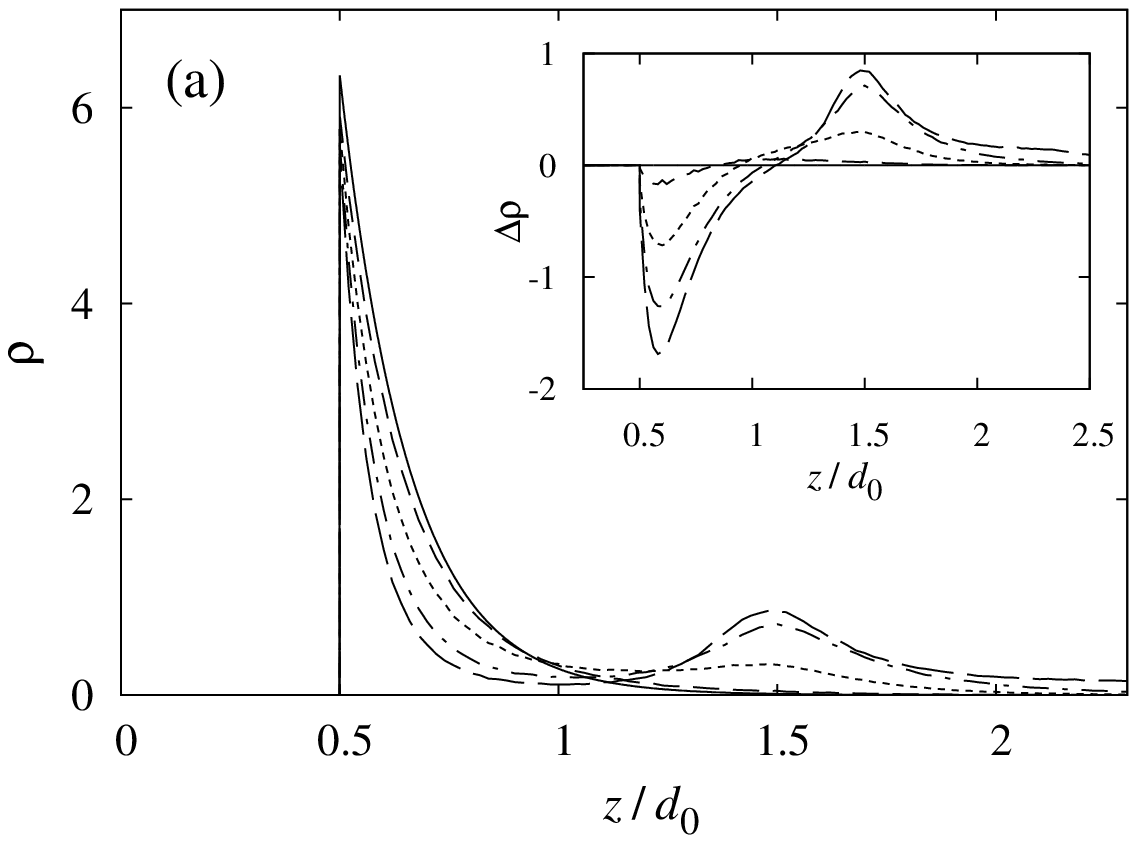}
\includegraphics[width=\ffraction\textwidth]{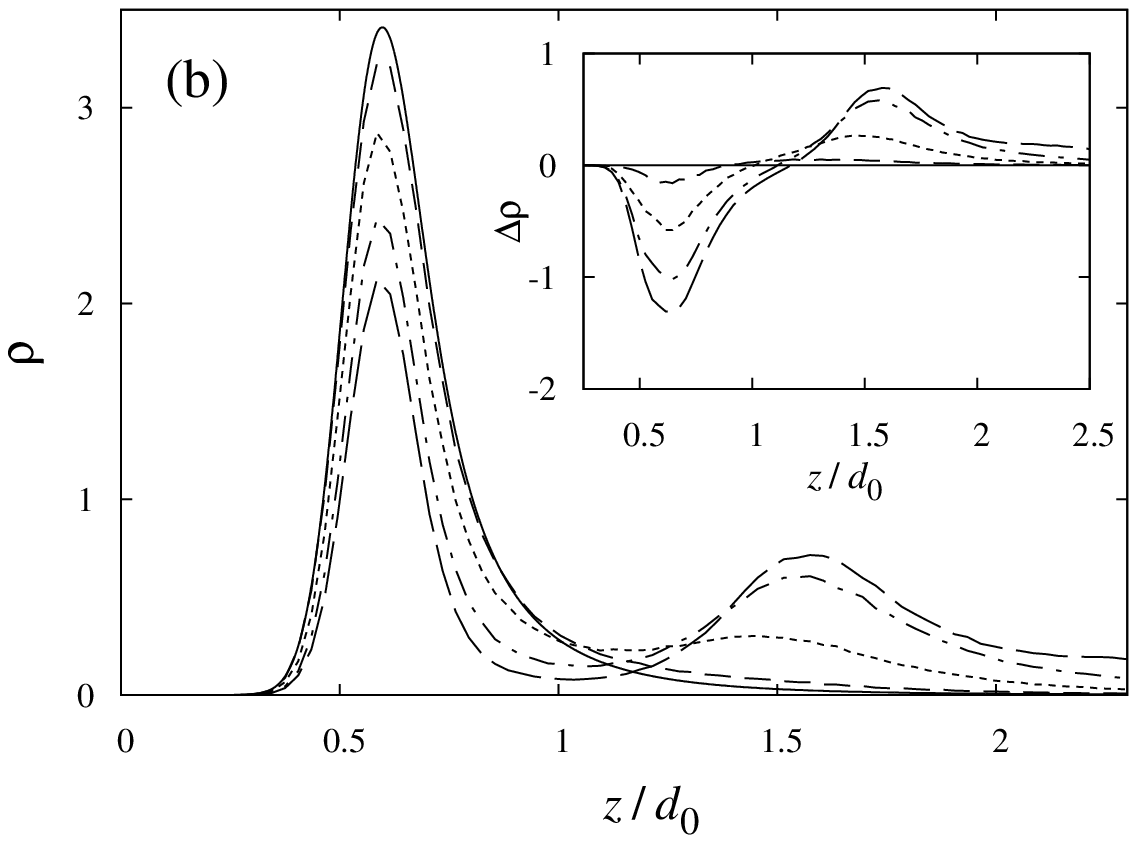}

\caption{Particle--wall distribution function for \subfig{a}
  monodisperse suspension; \subfig{b} polydisperse suspension with
  standard deviation of particle diameter
  $\sigma/d_0=0.15$. Simulation results for area fraction $\phi=0$
  (solid line), 0.3 (dashed), 0.6 (dotted), 0.9 (dot--dashed), 1.2
  (long-dashed). The insets show the deviation
  $\Delta\rho=\rho-\rho_{\phi=0}$ from the low-density distribution.
\label{particle-wall distributions simulations}}
\end{figure}

\begin{figure}
\includegraphics[width=\ffraction\textwidth]{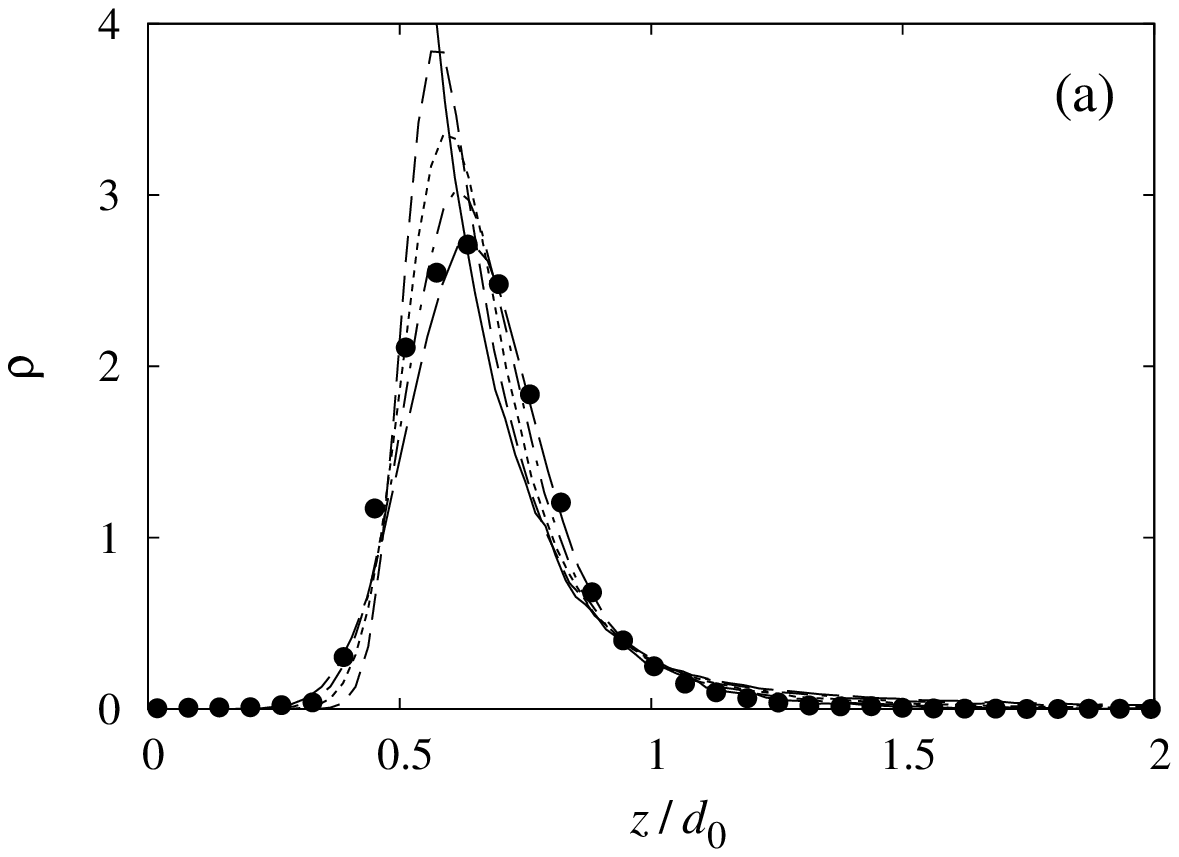}
\includegraphics[width=\ffraction\textwidth]{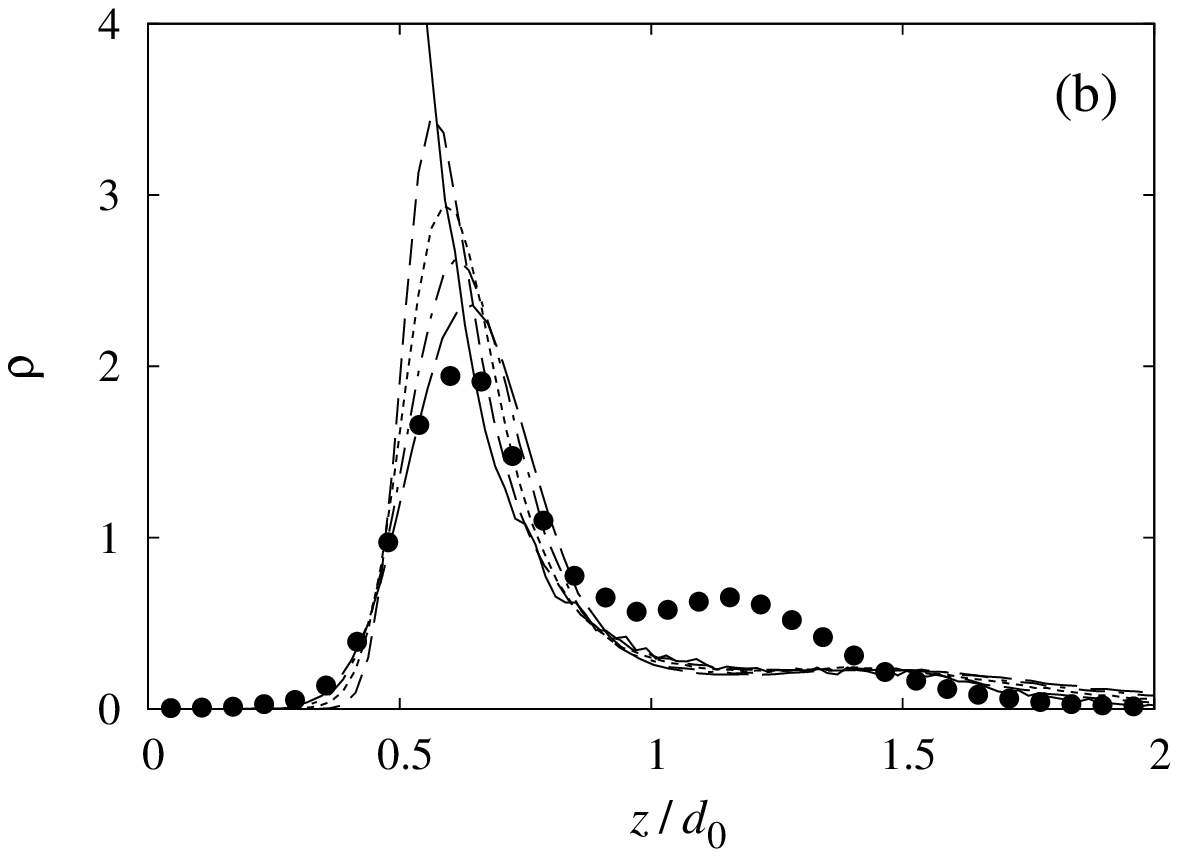}
\caption{Particle--wall distribution function for area fractions
  \subfig{a} $\phi=0.054$ and \subfig{b} $0.54$.  Experimental results
  (solid circles); simulation results for standard deviation of
  particle diameter $\sigma/d_0=0$ (solid line), $0.1$ (dashed),
  $0.15$ (dotted), $0.20$ (dashed--dotted), and $0.25$
  (long-dashed).  \label{particle-wall distributions: comparison with
    experiments}}
\end{figure}

\begin{figure}

\includegraphics[width=\ffraction\textwidth]{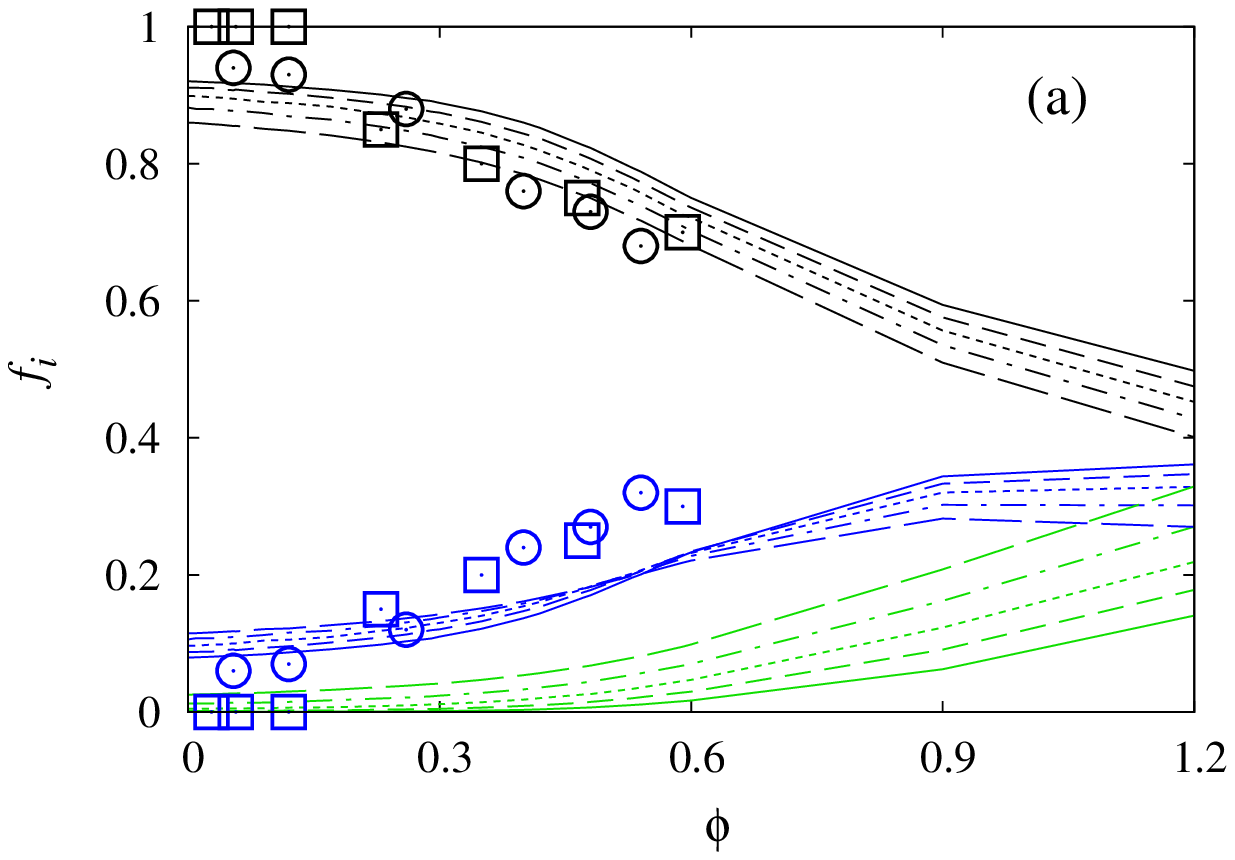}
\includegraphics[width=\ffraction\textwidth]{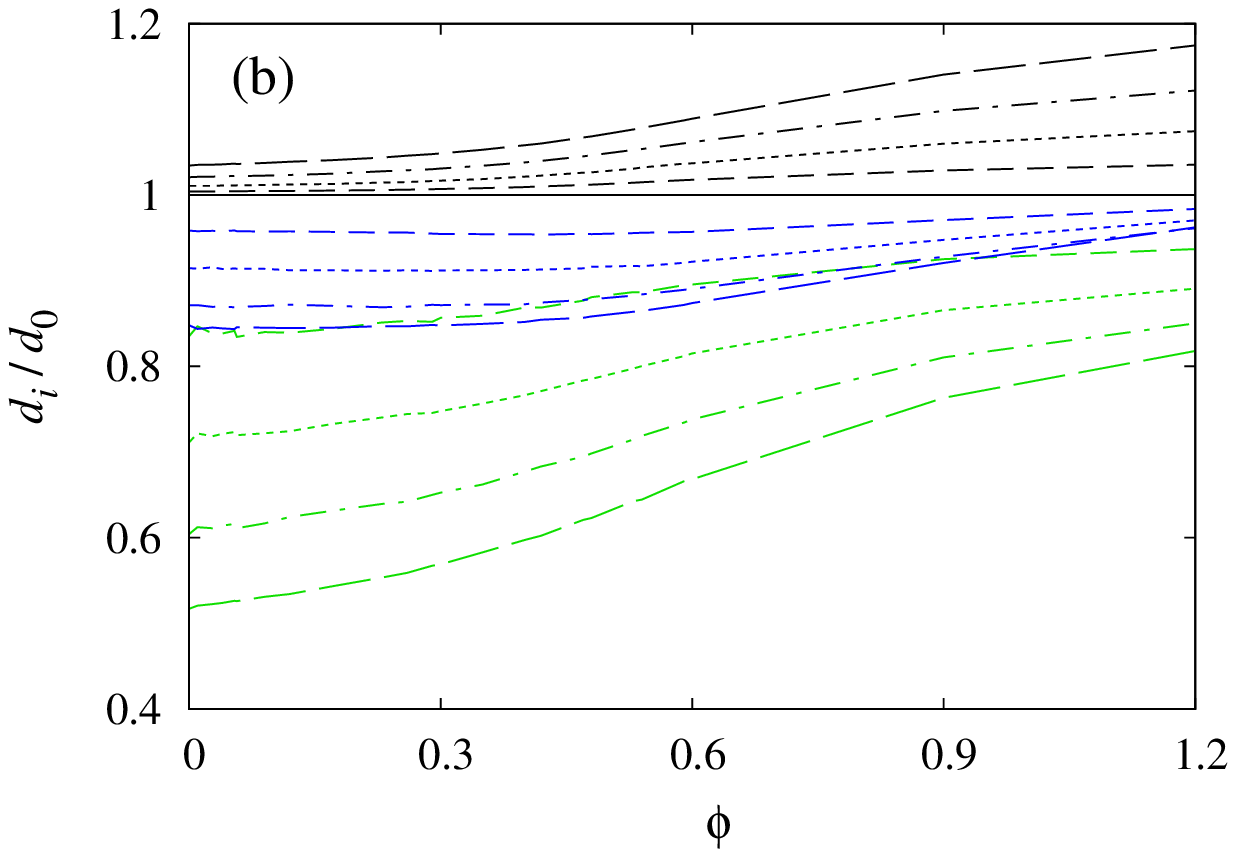}

\caption{\subfig{a} Occupation fraction $\occupationFraction{i}$ and
  \subfig{b} normalized average particle diameter $d_i$ in the first
  particle layer (black), second layer (blue), and third layer
  (green), vs the total area fraction $\phi$.  The results for a
  monodisperse system (solid lines) and polydisperse systems with
  standard deviation of particle diameter $\sigma/d_0=0.1$ (dashed),
  $0.15$ (dotted) $0.20$ (dash--dotted), and $0.25$ (long-dashed).
  The symbols represent experimental results from confocal imaging
  (circles) and 2D images (squares) for a suspension with salt
  concentration $\KCl=0.01\Mol$. Note that the experimental second
  layer corresponds to the sum of the second and third layers in the
  MC simulations.  The layer boundaries in the numerical calculations
  are set at $z_1=0.9d_0$, and $z_2=1.8d_0$ and in the experiments are
  obtained from Gaussian fitting (see Fig.\ \ref{fig:zhist}).
  \label{occupation numbers}}
  
\end{figure}

\begin{figure}

\includegraphics[width=\ffraction\textwidth]{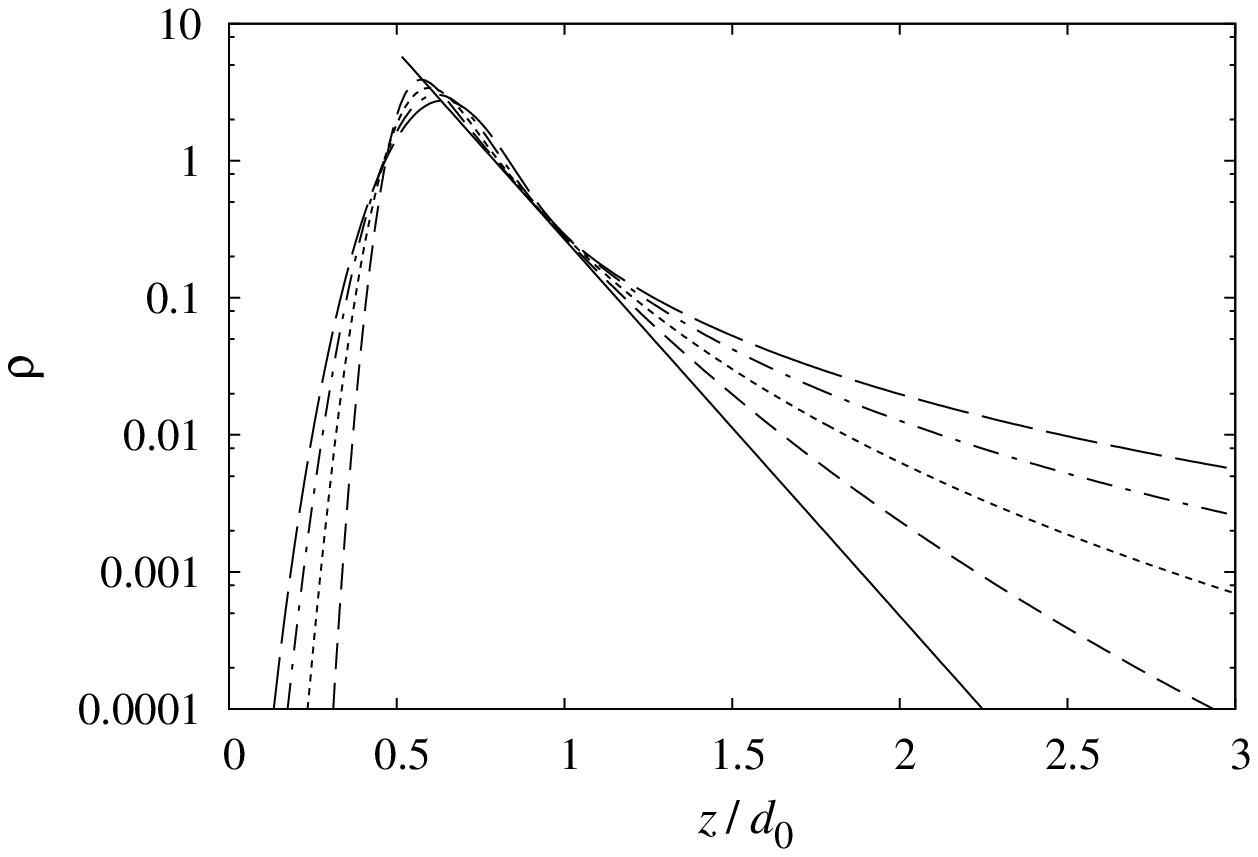}

\caption{Low-density limit of the near-wall particle distribution for
  a monodisperse system (solid line)  and polydisperse systems with 
  standard deviation of particle diameter $\sigma/d_0=0.1$ (dashed),
  $0.15$ (dotted) $0.20$ (dash--dotted), and $0.25$ (long-dashed). 
  \label{low-density distribution}}
\end{figure}

Here we present results of MC simulations of the equilibrium
microstructure of a HS suspension in the near wall region.  The HS
potential corresponds to the system with $\KCl=0.01\Mol$, for which
the electrostatic repulsion is negligible.  Since the suspension used
in the experiments is polydisperse, we consider both monodisperse and
polydisperse systems.

\subsubsection{Near-wall particle distribution}

Figure \ref{particle-wall distributions simulations} shows MC results
for the suspension density profile $\particleDistribution(z)$ for a
monodisperse suspension and polydisperse suspensions at several area
fractions.  Similarly to the experimental results, the simulations
show that there is a single layer of sedimented particles at low area
fractions $\phi$, and a two-layer microstructure at higher area
fractions. (Development of a third layer for $\phi\gtrsim0.9$ is
also noticeable in the region $z/d_0 \gtrsim 2$.)  Suspension
polydispersity results in broadening of the peaks of the particle
distribution.

A direct comparison between the experimental and simulation results is
presented in Fig.\ \ref{particle-wall distributions: comparison with
  experiments} for two values of the area fraction $\phi$.  At low
area fractions [Fig.\ \ref{particle-wall distributions: comparison
    with experiments}\subfig{a}] the agreement between the experiments
and simulations is good.  (The standard deviation of the particle-size
distribution for which the simulations match the experimental data,
$\sigma/d_0\approx 0.25$, is larger than the estimated standard
deviation $0.1<\sigma/d_0<0.15$ based on the manufacturer's
specifications; the additional spread of the experimentally observed
peak can be attributed to random errors of the particle height
evaluation from the confocal-microscopy images.)

A comparison of the numerical and experimental results at a higher
area fraction, as shown in Fig.\ \ref{particle-wall distributions:
  comparison with experiments}\subfig{b} [also see
  Figs.\ \ref{fig:zhist} and \ref{particle-wall distributions
    simulations}], reveals that (\textit{i}) the experimentally
observed second maximum of the density distribution develops at lower
area fractions than the corresponding maximum in the numerical
simulations; (\textit{ii}) the experimental second peak is narrower,
and its position is shifted towards the wall.  In contrast, the plots
of the excess distribution $\Delta\rho$ with respect to the
low-density limit, shown in Fig.\ \ref{fig:onset}\subfig{a} and the
insets of Fig.\ \ref{particle-wall distributions simulations},
indicate that the onset of the formation of the second layer occurs at
approximately the same area fraction according to the simulations and
experiments.  Moreover, the measured and calculated occupation
fractions of the layers are similar for all area fractions, as
depicted in Fig.\ \ref{occupation numbers}\subfig{a}.  A possible
source of the observed discrepancies between the experimental and
numerical results for the particle distribution $\rho(z)$ is described
in Appendix \ref{appendix}.  It also provides a plausible explanation
of the fact that the agreement between the experiments and MC
simulations for the layer occupation fractions $f_i$ is quite good in
spite of the discrepancies for $\particleDistribution(z)$.

\subsubsection{Polydispersity effects}

The results in Fig.\ \ref{occupation numbers}\subfig{a} show that the
occupation fraction of the first two layers is relatively insensitive
to the suspension polydispersity; in contrast, the occupation fraction
of the third layer strongly increases with the standard deviation of
particle diameter.  This increase stems from the presence of smaller
(lighter) particles in polydisperse systems: smaller particles tend to
migrate into the top layer, as evident from Fig.\ \ref{occupation
  numbers}\subfig{b}.  For dilute suspensions, the particle-size
segregation results in variation of the slope of $\log
\particleDistribution(z)$ with the distance from the wall, as
illustrated in Fig.\ \ref{low-density distribution}.  We estimate that
this variation causes an approximately 20\,\% uncertainty of the
calibration of the confocal height measurements described in
Sec.\ \ref{Height calibration}.

\subsection{A quasi-2D model of the equilibrium layered microstructure}
\label{A quasi-2D model of equilibrium layered microstructure}

Here we present a semi-quantitative theoretical model for evaluating
the occupation fractions $\occupationFraction{i}$ of the particle layers
in a sedimented colloidal suspension.  Our theory is based on the
assumption that the suspension microstructure can be approximated as a
collection of weakly coupled quasi-2D layers in
thermodynamic equilibrium with respect to particle exchange.  

The equilibrium condition for layers $i$ and $i+1$ is
\begin{equation}
\label{equilibrium between layers}
\mu_i+mgz_i=\mu_{i+1}+mgz_{i+1},
\end{equation}
where $\mu_i$ is the chemical potential of layer $i$, and $z_i$ is its
position.  In our model, $\mu_i$ is approximated as the chemical
potential of a 2D hard-disk fluid of area fraction $\phi_i$.  All disk
diameters are equal to the sphere diameter $d$, which corresponds to a
layer of spheres with the same vertical position $z$.

In the low area-fraction limit, the chemical potential of a hard-disk
fluid is
\begin{equation}
\label{chemical potential at low densities}
\mu_i=\kT\ln\phi_i+C(T),
\end{equation}
where $C(T)$ depends only on the temperature $T$.  According to the
equilibrium condition \eqref{equilibrium between layers} and equation
of state \eqref{chemical potential at low densities}, we thus have
\begin{equation}
\label{area fraction in the next layer}
\phi_{i+1}=r\phi_i,\qquad i=1,2,\ldots
\end{equation}
with the ratio $r$  given by the Boltzmann factor
\begin{equation}
\label{occupancy ratio}
r=\textrm{e}^{-\Deltaz/l},
\end{equation}
where $l$ is defined by Eq.\ \eqref{ldef} and $\Deltaz=z_{i+1}-z_i$.  We
assume that the layer separation $\Deltaz$ is independent of $i$.

For finite area fractions, relation \eqref{area fraction in the next
  layer} is replaced with 
\begin{equation}
\label{iteration for phi_i}
\phi_{i+1}=r(\phi_i)\phi_i,\qquad i=1,2\ldots,
\end{equation}
where the layer occupation ratio $r$ depends on the area fraction in
the adjacent layers.  The factor $r(\phi)$ is determined from the
equilibrium condition \eqref{equilibrium between layers} with the help
of the Gibbs--Duhem relation 
\begin{equation}
\label{Gibbs-Duhem}
\textrm{d}\mu=\frac{\pi d^2}{4}\phi^{-1}\textrm{d} p,
\end{equation}
where $p$ is the lateral 2D pressure within the layer.  Combining
\eqref{equilibrium between layers}, \eqref{iteration for phi_i}, and
\eqref{Gibbs-Duhem} yields
\begin{equation}
\label{differential equation for r}
\frac{\textrm{d}r}{\textrm{d}\phi}=
   r\left[\frac{p'(\phi)}{p'(r\phi)}-1\right],
\end{equation}
where
\begin{equation}
\label{p'}
p'=\left(\frac{\partial p}{\partial\phi}\right)_T.
\end{equation}
The differential equation \eqref{differential equation for r} is
solved for $r=r(\phi)$ with the boundary condition \eqref{occupancy
  ratio} at $\phi=0$.  Occupation fractions
$\occupationFraction{i}=\phi_i/\phi$ are then determined by iteration,
applying Eq.\ \eqref{iteration for phi_i} and the relation
\begin{equation}
\label{total area fraction on terms of partial fractions}
\phi=\sum_{i=1}^\infty\phi_i.
\end{equation}

\begin{figure}

\includegraphics[width=\ffraction\textwidth]{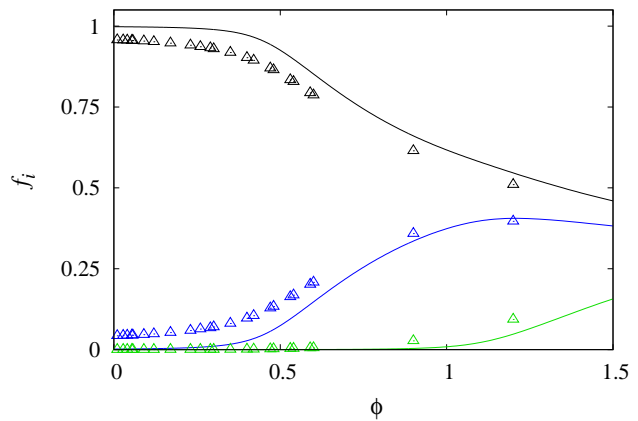}

\caption{Occupation fraction $f_i$ for the first (black), second
  (blue) and third (green) layer vs the total area fraction $\phi$ for
  a monodisperse HS suspension.  Theoretical results (solid
  lines); MC simulations (symbols).
  \label{quasi 2D model}}
\end{figure}

We have solved Eq.\ \eqref{differential equation for r} and determined
the occupation fractions $\occupationFraction{i}$ using the
scaled-particle-theory equation of state for hard disks
\cite{Helfand-Frisch-Lebowitz:1961}, 
\begin{equation}
\label{SPT pressure}
\frac{\pi d^2 p}{4\kT}=\frac{\phi}{(1-\phi)^2}.  
\end{equation}
The results of our calculations are presented in Fig.\ \ref{quasi 2D
  model} for a HS system with the same value of the sedimentation
length \eqref{value of dimensionless sedimentation constant} as in our
MC simulations.  Based on the separation between the first and second
peak of the suspension density profile shown in
Fig.\ \ref{particle-wall distributions simulations}\subfig{a}, the
calculations were performed for $\Deltaz/d=1$.

The theoretical results in Fig.\ \ref{quasi 2D model} are compared with
the MC simulations of a monodisperse HS suspension with the boundaries
between the layers set to $z_1=d$ and $z_2=2d$, consistent with the
peak positions.  The agreement between our simple theory and
simulations is quite good.  A similar agreement was obtained for other
values of the dimensional parameter $l/d$ (results not shown).

The layer boundaries used in Sec.\ \ref{Structure - numerical
  simulations} and \ref{Dynamics - numerical simulations} to compare
the MC results with experiments differ from the boundaries used in the
above model by approximately 10\,\%.  Due to the observed deviation
between the measured and simulated particle distributions (see
Fig.\ \ref{particle-wall distributions: comparison with experiments}),
it is not possible to define the layer boundaries in a unique,
equivalent way for the experimental and simulated systems.  Therefore,
the layer boundaries $z_1=0.9d$ and $z_2=1.8d$ used in
Sec.\ \ref{Structure - numerical simulations} and \ref{Dynamics -
  numerical simulations} were chosen based on the comparison between
the experimental and numerical results for the occupation fractions
and self-diffusivities in particle layers. 

\section{Particle dynamics}
\label{sec:dynamics}

\begin{figure}
\centering
\includegraphics[clip=true,scale=0.5]{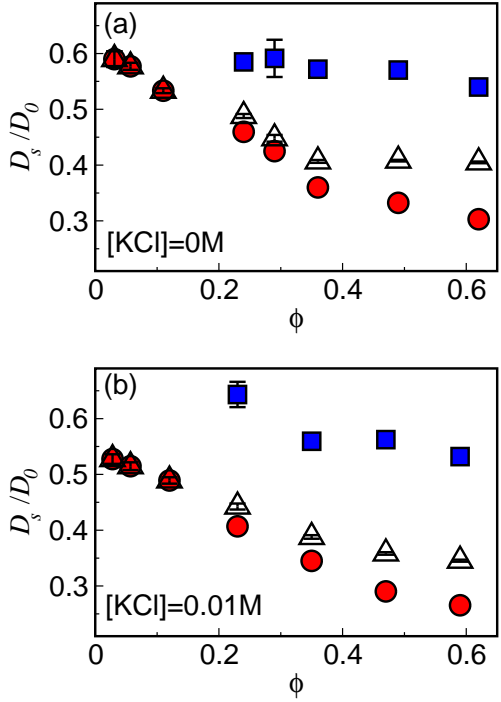}
\caption{Short-time self-diffusion coefficient $\Dself$, normalized by
  the Stokes-Einstein diffusion coefficient $D_0$, as a
  function of the total area fraction $\phi$; circles (squares) correspond
  to the first (second) layer, and triangles to the effective $\Dself$
  calculated by the weighted average of the self-diffusion coefficients for the two individual layers. (a) Suspension with no added salt. (b)
  Suspension with salt concentration $\KCl=0.01\Mol$.  }
\label{fig:def_gaus}
\end{figure}

\subsection{Experimental results}
\label{Dynamics - experimental results}

The short time self-diffusion coefficient in the $x$--$y$ plane,
$\Dself$, is determined for different total area fractions of the
sedimented particles by extracting the mean square displacement
\eqref{self diffusivity from mean-square displacement} from 2D
epifluorescent images of the first and second particle layer.  The
mean-square displacement is measured over a time interval $\tau$ that
is small compared to the structural relaxation time of the suspension,
to ensure that the measurements yield the short-time self-diffusion
coefficient.

The results are shown in Fig.~\ref{fig:def_gaus} for suspensions with
salt concentration $\KCl=0.01\Mol$ and salt-free suspensions with
$\KCl=0\Mol$.  The self-diffusion coefficient is expected to decrease
as the particle concentration increases; indeed, we observe this
decrease for both salt concentrations and in both layers, for
$\phi<0.4$.  In the case of $\KCl=0.01\Mol$, corresponding to $\lambda
= 5$ nm, the particles can get much closer to the cell floor, which in
turn results in lower values of the self-diffusion coefficient
compared to suspensions with $\KCl=0\Mol$.

Using a linear fit to the values of $\Dself/D_0$ for the low area
fractions, where there is no observable second layer, we can
extrapolate to $\phi=0$ and extract the self-diffusivity of a single
particle.  The extrapolated results agree well with the measurements
at very low concentrations $\phi<0.003$, as discussed in Sec.\ \ref{Mean
  particle height at low area fractions}.

From the known occupation fractions $\occupationFraction{1}$ and
$\occupationFraction{2}$ for each $\phi$ we can weigh the contribution
of each layer to the total self-diffusivity, and construct an
effective $\Dself$ of the whole suspension
(Fig.~\ref{fig:def_gaus}). As expected, for $\phi<0.4$ the effective
self-diffusion coefficient $\Dself$ decreases as $\phi$ is increased
in both salt concentrations. For larger $\phi$ we observe a flattening
of $\Dself$, which clearly indicates that the second layer becomes
dominant in those area fractions. This observation is supported also
by the saturation of $\phi_1$ at $\phi>0.45$
[Fig.~\ref{fig:occ}\subfig{b}].

\subsection{Numerical simulations}
\label{Dynamics - numerical simulations}

\begin{figure}[b]
\includegraphics[width=\ffraction\textwidth]{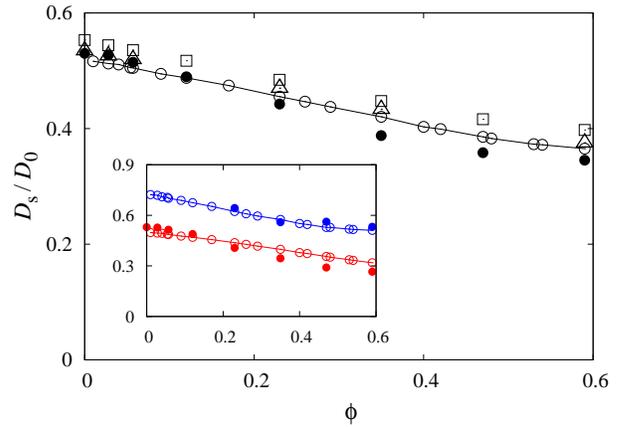}
\caption{Normalized short-time self-diffusion coefficient
  $\Dself/D_0$, as a function of the area fraction $\phi$ for a
  suspension with salt concentration $\KCl=0.01\Mol$.  The main panel
  shows $\Dself$ averaged over the whole system, and the inset shows
  $\Dself$ for the first (bottom, red) and second (top, blue) particle
  layer.  Experimental results (solid circles); simulation results
  (open symbols) for a monodisperse system (circles) and polydisperse
  systems with the standard deviation of the particle diameter
  $\sigma=0.1d_0$ (triangles) and $0.15d_0$ (squares).  Note that at
  low area fractions the triangles overlap with the solid circles.
  The lines are a guide for the eye. }\label{self-diffusion with salt}
\end{figure}

\begin{figure}
\includegraphics[width=\ffraction\textwidth]{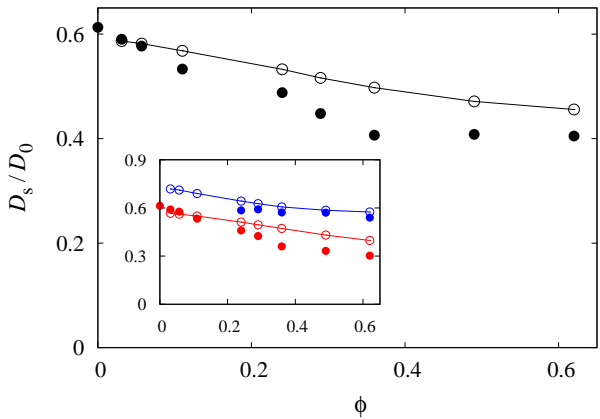}
\caption{Normalized short-time self-diffusion coefficient
  $\Dself/D_0$, as a function of the area fraction $\phi$ for a
  suspension with no salt.  Symbols are the same as in
  Fig.\ \ref{self-diffusion with salt}.  Results are shown only for monodisperse suspensions. \label{self-diffusion with no salt}}
\end{figure}

\begin{figure}
\includegraphics[width=\ffraction\textwidth]{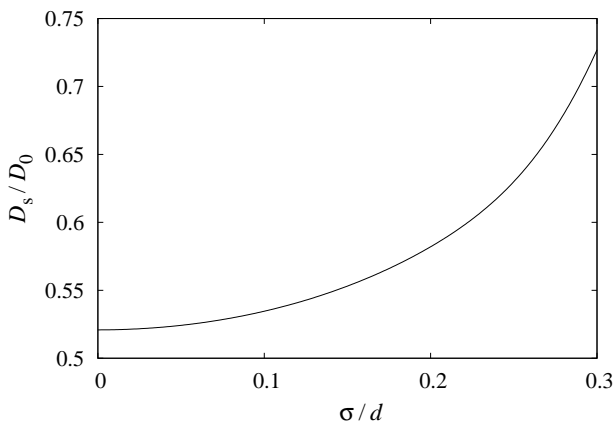}
\caption{Normalized short-time self-diffusion coefficient $\Dself/D_0$
  in the low-area-fraction limit $\phi=0$ (for the system with salt) as a function of the
  suspension polydispersity.  \label{self-diffusion low area
    fraction}}
\end{figure}

The results of our numerical simulations for the short-time lateral
self-diffusion coefficient $\Dself$ in a HS system are
presented in Fig.\ \ref{self-diffusion with salt} for a monodisperse
suspension and for polydisperse suspensions with $\sigma/d_0=0.1$ and
$0.15$.  Figure \ref{self-diffusion with no salt} shows the
corresponding results for a system of monodisperse hard spheres with
particle--wall and particle--particle electrostatic repulsion
\eqref{V} and \eqref{V'}.

The results depicted in Fig.\ \ref{self-diffusion with salt}
indicate that for moderately polydisperse suspensions (in the range
corresponding to the polydispersity of silica particles used in the
experiments), the self-diffusion coefficient is only moderately
dependent on $\sigma/d_0$.  For larger values of the variance of
particle diameters, the normalized self-diffusion coefficient
$\Dself/D_0$ significantly increases with the degree of the
polydispersity, because the mobility is dominated by small particles.
This increase is illustrated in Fig.\ \ref{self-diffusion low area
  fraction} for a suspension in the low-area-fraction limit $\phi=0$.

The results of our hydrodynamic calculations for a HS suspension and
for a suspension with screened electrostatic repulsion are compared
with experimental results for suspensions with $\KCl=0.01\Mol$
(Fig.\ \ref{self-diffusion with salt}) and $\KCl=0\Mol$
(Fig.\ \ref{self-diffusion with no salt}).  For the system with salt,
the measured values are slightly closer to their numerical analogs
than in the absence of salt. Summarizing, the experimental and
numerical results agree well both for the overall self-diffusivity and
for the self-diffusivity in individual particle layers.
\section{Discussion}
\label{sec:discuss}

In this paper we have studied in detail the structure and dynamics of
quasi-2D colloidal suspensions near a wall, comparing experiment and
theory. Our central result is a rather sharp formation of a distinct
second layer at an area fraction of $\phi\sim0.3$. This value is much
lower than the area fraction required for close-packing or other 2D
structural changes such as the formation of hexatic or crystalline
order. One important consequence of this result concerns the apparent
self-diffusion of the particles in the suspension and its dependence
on particle density. Due to the higher mobility of the particles in
the elevated layer, the effective diffusivity is higher and levels off
as particle density increases. 
The experimentally observed behavior could be interpreted incorrectly if one is unaware of the layering (or stratifying) effect.

We find good agreement between experimental and simulation results for
the occupation fractions of the first and second layers and for the
lateral self-diffusivity (both for the entire suspension and in the
individual layers).  However, we also find an unexpected discrepancy
in the position and the height of the second peak in the near-wall
particle distribution. While the source of this discrepancy is
unknown, one possibility, related to optical aberrations, is suggested
in Appendix \ref{appendix}.  On the other hand, the difference between
theory and experiment might also be a result of an actual physical
effect, such as more complicated electrostatic interactions setting in
at higher layer densities.

Another new insight put forth in this study is the significant effect
that polydispersity has on the occupation and composition of layers
close to the bottom wall, even in the case of a relatively small
dispersion of particle sizes. The effect of polydispersity is evident
already at low densities, since the smaller and larger particles
segregate into the upper and lower layers, respectively.  We expect
the phenomena described here to be quite general and to be manifested
in any such system where the sedimentation length $l$ is of the order
of the particle diameter.  This conclusion is supported by the
appearance of the phenomena both in experiments and in Monte--Carlo
simulations.

An important outcome of this paper is the construction of a very
simple theoretical quasi-2D model of the layered microstructure in
thermodynamic equilibrium. Such systems have been analyzed earlier
using density-functional theory \cite{Chen2006}, but our theoretical
model is much simpler and easier to apply. We have demonstrated that
the model approximates well the experimental and numerical results for
the system studied in this work.

We conclude with three open issues. Layering phenomena near a wall are
well documented in 3D suspensions as well
\cite{VanWinkle1988,GonzalezMozuelos1991,Zurita_Gotor-Blawzdziewicz-Wajnryb:2012}. An
interesting question is whether this perturbation to the 3D pair
correlation function could be fundamentally related to the sequential
layering reported here. The structural features near the wall should
also affect two- and many-particle dynamics in the quasi-2D
suspensions, which can be characterized by two-point
microrheology. Finally, taking a more detailed account of
interparticle forces such as strong electrostatic interactions may
hopefully provide deeper understanding of the effects observed in this
work.

\acknowledgments H.D. wishes to thank the Polish Academy of Sciences
for its hospitality. This research has been supported by the Israel
Science Foundation (Grants No.\ 8/10 and No.\ 164/14) and by the Marie
Curie Reintegration Grant (PIRG04-GA-2008-239378). A.S.--S acknowledges
funding from the Tel-Aviv University Center for Nanoscience and
Nanotechnology. M.L.E.--J. and E.W. were supported in part by
Narodowe Centrum Nauki (National Science Centre) under grant
No. 2012/05/B/ST8/03010.  J.B. would like to acknowledge the financial
support from National Science Foundation (NSF) Grant No. CBET 1059745.

\appendix

\section{}
\label{appendix}

\begin{figure}[t]
\includegraphics[width=\ffraction\textwidth]{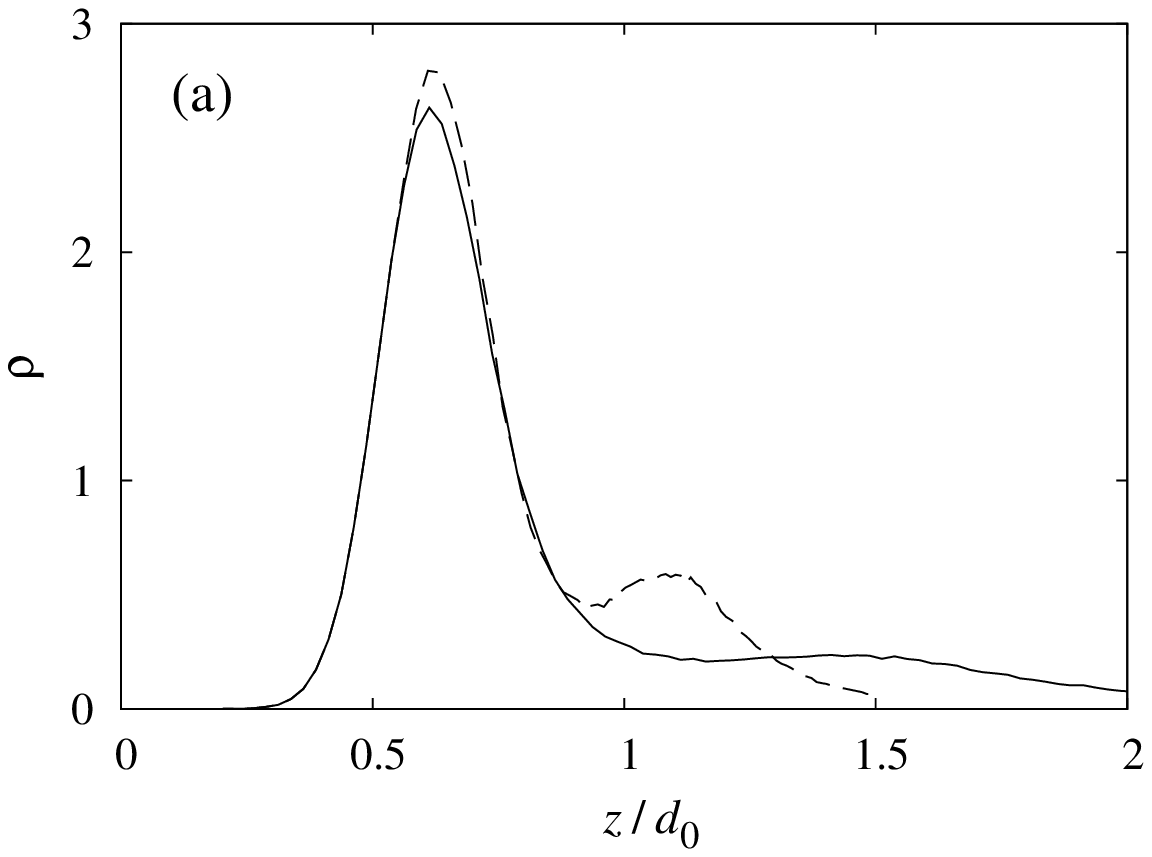}
\includegraphics[width=\ffraction\textwidth]{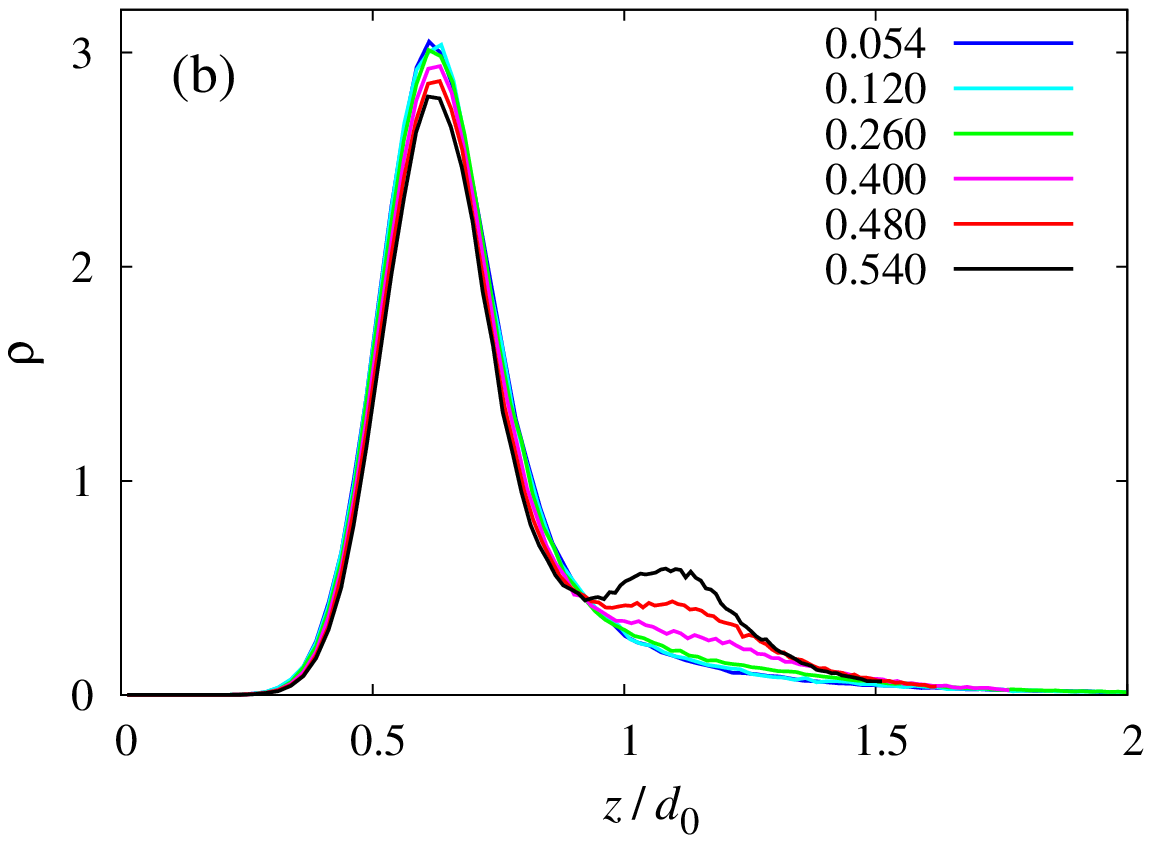}
\caption{The effect of the model distortion transformation
  \eqref{distortion model} on the particle distribution $\rho(z)$ in a
  HS suspension with the standard deviation of the particle
  distribution $\sigma/d_0=0.2$.  \subfig{a} A comparison of the MC
  result for $\rho(z)$ (solid line) with the transformed distribution
  \eqref{rescaled probability} (dashed line) at the area fraction
  $\phi=0.54$; \subfig{b} the transformed distribution for different
  area fractions (as labeled).  The parameters of the transformation
  \eqref{distortion model} are $z_1/d=0.55$, $z_2/d=1.2$, and
  $\alpha=2.5$.  \label{Distorted distribution}}
\end{figure}

We present a simple model to support a hypothesis that the
discrepancy between the measured and calculated near-wall particle
distributions stems from optical aberration caused by nonuniform
optical properties of the suspension in the near-wall region.  We
assume that such aberration produces a nonlinear rescaling of the
coordinate $z$,
\begin{equation}
\label{rescaling of z}
\tilde z=\tilde z(z),
\end{equation}
where $z$ is the actual and $\tilde z$ is the measured particle
position.  The rescaling \eqref{rescaling of z} results in the
corresponding transformation of the particle density
\begin{equation}
\label{rescaled probability}
\tilde\particleDistribution(\tilde z)=\particleDistribution(z)\frac{d\,z}{d\,\tilde z}.
\end{equation}
To demonstrate that a distortion \eqref{rescaling of z} can produce
the observed shift and change of height of the features of the
distribution $\particleDistribution$, we consider an \textit{ad hoc}
distortion model with the transformation between the measured and
actual vertical coordinates given by the equations
\begin{subequations}
\label{distortion model}
\begin{equation}
\label{Jacobian}
\frac{d\,\tilde z}{d\,z}=\left\{\begin{array}{ll}
1; &z<z_1,\\
1-(1-b)\displaystyle\frac{z-z_1}{z_2-z_1};\quad\quad &z_1\le z\le z_2,\\
b; &z_2<z,
\end{array}
\right.
\end{equation}
and
\begin{equation}
\label{expression for b}
b=\alpha\phi,
\end{equation}
\end{subequations}
where $z_1$, $z_2$, and $\alpha$ are the model parameters.  The
transformation \eqref{distortion model} describes position-dependent
coordinate contraction with the amplitude gradually increasing in the
region $z_1\le z\le z_2$ (the region where the second peak occurs
according to the experimental data).  The overall deviation of the
Jacobian \eqref{distortion model} from unity is proportional to the
area fraction of the suspension layer.

Figure \ref{Distorted distribution}\subfig{a} compares the distorted
distribution \eqref{rescaled probability} with the corresponding
untransformed distribution $\particleDistribution(z)$ obtained from MC
simulations of a HS suspension at the area fraction $\phi=0.54$.
Figure \ref{Distorted distribution}\subfig{b} presents the distorted
distribution for the set of area fractions for which experimental
results are depicted in Fig.\ \ref{fig:zhist}\subfig{a}.  The
parameter values of the transformation \eqref{distortion model} are
given in the figure caption.

The results show that the coordinate transformation \eqref{Jacobian}
shifts the position of the second particle layer to the left and
produces a corresponding enhancement of the peak of particle
distribution, similar to the experimentally observed features of the
distributions depicted in Figs.\ \ref{fig:zhist}\subfig{a} and
\ref{particle-wall distributions: comparison with
  experiments}\subfig{b}.  Thus our calculations provide indirect
support to our optical-distortion hypothesis.  The distortion hypothesis
can also explain why the measured and calculated occupation fractions
and self-diffusivities of the particles in the top layer agree well
(see Figs.\ \ref{occupation numbers} and \ref{self-diffusion with
  salt}), in spite of the fact that the observed and calculated
positions of the layer differ significantly.

It is an open question what the source of the distortion
\eqref{rescaling of z} might be.  Since the suspension is imaged from
above in our confocal-microscopy system, we hypothesize that
reflection of laser light from the first (bottom) particle layer
results in stray illumination of the second layer, producing distorted
particle height measurements.  The optical distortion hypothesis can
be verified by experiments using refractive-index matched suspensions,
but such investigations are beyond the scope of the present study.


\begin{thebibliography}{52}%
\makeatletter
\providecommand \@ifxundefined [1]{%
 \@ifx{#1\undefined}
}%
\providecommand \@ifnum [1]{%
 \ifnum #1\expandafter \@firstoftwo
 \else \expandafter \@secondoftwo
 \fi
}%
\providecommand \@ifx [1]{%
 \ifx #1\expandafter \@firstoftwo
 \else \expandafter \@secondoftwo
 \fi
}%
\providecommand \natexlab [1]{#1}%
\providecommand \enquote  [1]{``#1''}%
\providecommand \bibnamefont  [1]{#1}%
\providecommand \bibfnamefont [1]{#1}%
\providecommand \citenamefont [1]{#1}%
\providecommand \href@noop [0]{\@secondoftwo}%
\providecommand \href [0]{\begingroup \@sanitize@url \@href}%
\providecommand \@href[1]{\@@startlink{#1}\@@href}%
\providecommand \@@href[1]{\endgroup#1\@@endlink}%
\providecommand \@sanitize@url [0]{\catcode `\\12\catcode `\$12\catcode
  `\&12\catcode `\#12\catcode `\^12\catcode `\_12\catcode `\%12\relax}%
\providecommand \@@startlink[1]{}%
\providecommand \@@endlink[0]{}%
\providecommand \url  [0]{\begingroup\@sanitize@url \@url }%
\providecommand \@url [1]{\endgroup\@href {#1}{\urlprefix }}%
\providecommand \urlprefix  [0]{URL }%
\providecommand \Eprint [0]{\href }%
\providecommand \doibase [0]{http://dx.doi.org/}%
\providecommand \selectlanguage [0]{\@gobble}%
\providecommand \bibinfo  [0]{\@secondoftwo}%
\providecommand \bibfield  [0]{\@secondoftwo}%
\providecommand \translation [1]{[#1]}%
\providecommand \BibitemOpen [0]{}%
\providecommand \bibitemStop [0]{}%
\providecommand \bibitemNoStop [0]{.\EOS\space}%
\providecommand \EOS [0]{\spacefactor3000\relax}%
\providecommand \BibitemShut  [1]{\csname bibitem#1\endcsname}%
\let\auto@bib@innerbib\@empty
%</preamble>
\bibitem [{\citenamefont {Van~Winkle}\ and\ \citenamefont
  {Murray}(1988)}]{VanWinkle1988}%
  \BibitemOpen
  \bibfield  {author} {\bibinfo {author} {\bibfnamefont {D.~H.}\ \bibnamefont
  {Van~Winkle}}\ and\ \bibinfo {author} {\bibfnamefont {C.~A.}\ \bibnamefont
  {Murray}},\ }\href@noop {} {\bibfield  {journal} {\bibinfo  {journal} {J.
  Chem. Phys.}\ }\textbf {\bibinfo {volume} {89}},\ \bibinfo {pages} {3885}
  (\bibinfo {year} {1988})}\BibitemShut {NoStop}%
\bibitem [{\citenamefont {Gonz\'alez-Mozuelos}\ \emph
  {et~al.}(1991)\citenamefont {Gonz\'alez-Mozuelos}, \citenamefont
  {Medina-Noyola}, \citenamefont {D'Aguanno}, \citenamefont
  {M\'endez-Alcaraz},\ and\ \citenamefont {Klein}}]{GonzalezMozuelos1991}%
  \BibitemOpen
  \bibfield  {author} {\bibinfo {author} {\bibfnamefont {P.}~\bibnamefont
  {Gonz\'alez-Mozuelos}}, \bibinfo {author} {\bibfnamefont {M.}~\bibnamefont
  {Medina-Noyola}}, \bibinfo {author} {\bibfnamefont {B.}~\bibnamefont
  {D'Aguanno}}, \bibinfo {author} {\bibfnamefont {J.~M.}\ \bibnamefont
  {M\'endez-Alcaraz}}, \ and\ \bibinfo {author} {\bibfnamefont
  {R.}~\bibnamefont {Klein}},\ }\href@noop {} {\bibfield  {journal} {\bibinfo
  {journal} {J. Chem. Phys.}\ }\textbf {\bibinfo {volume} {95}},\ \bibinfo
  {pages} {2006} (\bibinfo {year} {1991})}\BibitemShut {NoStop}%
\bibitem [{\citenamefont {Zurita-Gotor}\ \emph {et~al.}(2012)\citenamefont
  {Zurita-Gotor}, \citenamefont {B{\l}awzdziewicz},\ and\ \citenamefont
  {Wajnryb}}]{Zurita_Gotor-Blawzdziewicz-Wajnryb:2012}%
  \BibitemOpen
  \bibfield  {author} {\bibinfo {author} {\bibfnamefont {M.}~\bibnamefont
  {Zurita-Gotor}}, \bibinfo {author} {\bibfnamefont {J.}~\bibnamefont
  {B{\l}awzdziewicz}}, \ and\ \bibinfo {author} {\bibfnamefont
  {E.}~\bibnamefont {Wajnryb}},\ }\href@noop {} {\bibfield  {journal} {\bibinfo
   {journal} {Phys. Rev. Lett.}\ }\textbf {\bibinfo {volume} {108}},\ \bibinfo
  {pages} {068301} (\bibinfo {year} {2012})}\BibitemShut {NoStop}%
\bibitem [{\citenamefont {Happel}\ and\ \citenamefont
  {Brenner}(1983)}]{Happel}%
  \BibitemOpen
  \bibfield  {author} {\bibinfo {author} {\bibfnamefont {J.}~\bibnamefont
  {Happel}}\ and\ \bibinfo {author} {\bibfnamefont {H.}~\bibnamefont
  {Brenner}},\ }\href@noop {} {\emph {\bibinfo {title} {Low Reynolds Number
  Hydrodynamics}}}\ (\bibinfo  {publisher} {Martinus Nijhoff},\ \bibinfo
  {address} {The Hague},\ \bibinfo {year} {1983})\BibitemShut {NoStop}%
\bibitem [{\citenamefont {Perkins}\ and\ \citenamefont
  {Jones}(1992)}]{Perkins1992}%
  \BibitemOpen
  \bibfield  {author} {\bibinfo {author} {\bibfnamefont {G.~S.}\ \bibnamefont
  {Perkins}}\ and\ \bibinfo {author} {\bibfnamefont {R.~B.}\ \bibnamefont
  {Jones}},\ }\href@noop {} {\bibfield  {journal} {\bibinfo  {journal} {Physica
  A}\ }\textbf {\bibinfo {volume} {189}},\ \bibinfo {pages} {447} (\bibinfo
  {year} {1992})}\BibitemShut {NoStop}%
\bibitem [{\citenamefont {Cichocki}\ and\ \citenamefont
  {Jones}(1998)}]{Cichocki-Jones:1998}%
  \BibitemOpen
  \bibfield  {author} {\bibinfo {author} {\bibfnamefont {B.}~\bibnamefont
  {Cichocki}}\ and\ \bibinfo {author} {\bibfnamefont {R.~B.}\ \bibnamefont
  {Jones}},\ }\href@noop {} {\bibfield  {journal} {\bibinfo  {journal} {Physica
  A}\ }\textbf {\bibinfo {volume} {258}},\ \bibinfo {pages} {273} (\bibinfo
  {year} {1998})}\BibitemShut {NoStop}%
\bibitem [{\citenamefont {Walz}\ and\ \citenamefont
  {Suresh}(1995)}]{Walz-Suresh:1995}%
  \BibitemOpen
  \bibfield  {author} {\bibinfo {author} {\bibfnamefont {J.~Y.}\ \bibnamefont
  {Walz}}\ and\ \bibinfo {author} {\bibfnamefont {L.}~\bibnamefont {Suresh}},\
  }\href@noop {} {\bibfield  {journal} {\bibinfo  {journal} {J. Chem. Phys.}\
  }\textbf {\bibinfo {volume} {103}},\ \bibinfo {pages} {10714} (\bibinfo
  {year} {1995})}\BibitemShut {NoStop}%
\bibitem [{\citenamefont {Prieve}(1999)}]{Prieve1999}%
  \BibitemOpen
  \bibfield  {author} {\bibinfo {author} {\bibfnamefont {D.~C.}\ \bibnamefont
  {Prieve}},\ }\href@noop {} {\bibfield  {journal} {\bibinfo  {journal} {Adv.
  Colloid Interface Sci.}\ }\textbf {\bibinfo {volume} {82}},\ \bibinfo {pages}
  {93} (\bibinfo {year} {1999})}\BibitemShut {NoStop}%
\bibitem [{\citenamefont {Sholl}\ \emph {et~al.}(2000)\citenamefont {Sholl},
  \citenamefont {Fenwick}, \citenamefont {Atman},\ and\ \citenamefont
  {Prieve}}]{Prieve2000}%
  \BibitemOpen
  \bibfield  {author} {\bibinfo {author} {\bibfnamefont {D.~S.}\ \bibnamefont
  {Sholl}}, \bibinfo {author} {\bibfnamefont {M.~K.}\ \bibnamefont {Fenwick}},
  \bibinfo {author} {\bibfnamefont {E.}~\bibnamefont {Atman}}, \ and\ \bibinfo
  {author} {\bibfnamefont {D.~C.}\ \bibnamefont {Prieve}},\ }\href@noop {}
  {\bibfield  {journal} {\bibinfo  {journal} {J. Chem. Phys.}\ }\textbf
  {\bibinfo {volume} {113}},\ \bibinfo {pages} {9268} (\bibinfo {year}
  {2000})}\BibitemShut {NoStop}%
\bibitem [{\citenamefont {Carbajal-Tinoco}\ \emph {et~al.}(2007)\citenamefont
  {Carbajal-Tinoco}, \citenamefont {Lopez-Fernandez},\ and\ \citenamefont
  {Arauz-Lara}}]{CarbajalTinoco2007}%
  \BibitemOpen
  \bibfield  {author} {\bibinfo {author} {\bibfnamefont {M.~D.}\ \bibnamefont
  {Carbajal-Tinoco}}, \bibinfo {author} {\bibfnamefont {R.}~\bibnamefont
  {Lopez-Fernandez}}, \ and\ \bibinfo {author} {\bibfnamefont {J.~L.}\
  \bibnamefont {Arauz-Lara}},\ }\href@noop {} {\bibfield  {journal} {\bibinfo
  {journal} {Phys. Rev. Lett.}\ }\textbf {\bibinfo {volume} {99}},\ \bibinfo
  {pages} {138303} (\bibinfo {year} {2007})}\BibitemShut {NoStop}%
\bibitem [{\citenamefont {Blawzdziewicz}\ \emph {et~al.}(2010)\citenamefont
  {Blawzdziewicz}, \citenamefont {Ekiel-Je\.{z}ewska},\ and\ \citenamefont
  {Wajnryb}}]{Blawzdziewicz2010}%
  \BibitemOpen
  \bibfield  {author} {\bibinfo {author} {\bibfnamefont {J.}~\bibnamefont
  {Blawzdziewicz}}, \bibinfo {author} {\bibfnamefont {M.~L.}\ \bibnamefont
  {Ekiel-Je\.{z}ewska}}, \ and\ \bibinfo {author} {\bibfnamefont
  {E.}~\bibnamefont {Wajnryb}},\ }\href@noop {} {\bibfield  {journal} {\bibinfo
   {journal} {J. Chem. Phys.}\ }\textbf {\bibinfo {volume} {133}},\ \bibinfo
  {pages} {114703} (\bibinfo {year} {2010})}\BibitemShut {NoStop}%
\bibitem [{\citenamefont {Dufresne}\ \emph {et~al.}(2000)\citenamefont
  {Dufresne}, \citenamefont {Squires}, \citenamefont {Brenner},\ and\
  \citenamefont {Grier}}]{Dufresne2000}%
  \BibitemOpen
  \bibfield  {author} {\bibinfo {author} {\bibfnamefont {E.~R.}\ \bibnamefont
  {Dufresne}}, \bibinfo {author} {\bibfnamefont {T.~M.}\ \bibnamefont
  {Squires}}, \bibinfo {author} {\bibfnamefont {M.~P.}\ \bibnamefont
  {Brenner}}, \ and\ \bibinfo {author} {\bibfnamefont {D.~G.}\ \bibnamefont
  {Grier}},\ }\href@noop {} {\bibfield  {journal} {\bibinfo  {journal} {Phys.
  Rev. Lett.}\ }\textbf {\bibinfo {volume} {85}},\ \bibinfo {pages} {3317}
  (\bibinfo {year} {2000})}\BibitemShut {NoStop}%
\bibitem [{\citenamefont {Cichocki}\ \emph {et~al.}(2007)\citenamefont
  {Cichocki}, \citenamefont {Ekiel-Je\.{z}ewska},\ and\ \citenamefont
  {Wajnryb}}]{Cichocki2007}%
  \BibitemOpen
  \bibfield  {author} {\bibinfo {author} {\bibfnamefont {B.}~\bibnamefont
  {Cichocki}}, \bibinfo {author} {\bibfnamefont {M.~L.}\ \bibnamefont
  {Ekiel-Je\.{z}ewska}}, \ and\ \bibinfo {author} {\bibfnamefont
  {E.}~\bibnamefont {Wajnryb}},\ }\href@noop {} {\bibfield  {journal} {\bibinfo
   {journal} {J. Chem. Phys.}\ }\textbf {\bibinfo {volume} {126}},\ \bibinfo
  {pages} {184704} (\bibinfo {year} {2007})}\BibitemShut {NoStop}%
\bibitem [{\citenamefont {Zurita-Gotor}\ \emph {et~al.}(2007)\citenamefont
  {Zurita-Gotor}, \citenamefont {B{\l}awzdziewicz},\ and\ \citenamefont
  {Wajnryb}}]{Zurita_Gotor-Blawzdziewicz-Wajnryb:2007b}%
  \BibitemOpen
  \bibfield  {author} {\bibinfo {author} {\bibfnamefont {M.}~\bibnamefont
  {Zurita-Gotor}}, \bibinfo {author} {\bibfnamefont {J.}~\bibnamefont
  {B{\l}awzdziewicz}}, \ and\ \bibinfo {author} {\bibfnamefont
  {E.}~\bibnamefont {Wajnryb}},\ }\href@noop {} {\bibfield  {journal} {\bibinfo
   {journal} {J. Fluid Mech.}\ }\textbf {\bibinfo {volume} {592}},\ \bibinfo
  {pages} {447} (\bibinfo {year} {2007})}\BibitemShut {NoStop}%
\bibitem [{\citenamefont {Anekal}\ and\ \citenamefont
  {Bevan}(2006)}]{Anekal2006}%
  \BibitemOpen
  \bibfield  {author} {\bibinfo {author} {\bibfnamefont {S.~G.}\ \bibnamefont
  {Anekal}}\ and\ \bibinfo {author} {\bibfnamefont {M.~A.}\ \bibnamefont
  {Bevan}},\ }\href@noop {} {\bibfield  {journal} {\bibinfo  {journal} {J.
  Chem. Phys.}\ }\textbf {\bibinfo {volume} {125}},\ \bibinfo {pages} {034906}
  (\bibinfo {year} {2006})}\BibitemShut {NoStop}%
\bibitem [{\citenamefont {Michailidou}\ \emph {et~al.}(2009)\citenamefont
  {Michailidou}, \citenamefont {Petekidis}, \citenamefont {Swan},\ and\
  \citenamefont {Brady}}]{Michailidou2009}%
  \BibitemOpen
  \bibfield  {author} {\bibinfo {author} {\bibfnamefont {V.~N.}\ \bibnamefont
  {Michailidou}}, \bibinfo {author} {\bibfnamefont {G.}~\bibnamefont
  {Petekidis}}, \bibinfo {author} {\bibfnamefont {J.~W.}\ \bibnamefont {Swan}},
  \ and\ \bibinfo {author} {\bibfnamefont {J.~F.}\ \bibnamefont {Brady}},\
  }\href@noop {} {\bibfield  {journal} {\bibinfo  {journal} {Phys. Rev. Lett.}\
  }\textbf {\bibinfo {volume} {102}},\ \bibinfo {pages} {068302} (\bibinfo
  {year} {2009})}\BibitemShut {NoStop}%
\bibitem [{\citenamefont {Cichocki}\ \emph {et~al.}(2010)\citenamefont
  {Cichocki}, \citenamefont {Wajnryb}, \citenamefont {B{\l}awzdziewicz},
  \citenamefont {Dhont},\ and\ \citenamefont
  {Lang}}]{Cichocki-Wajnryb-Blawzdziewicz-Dhont-Lang:2010}%
  \BibitemOpen
  \bibfield  {author} {\bibinfo {author} {\bibfnamefont {B.}~\bibnamefont
  {Cichocki}}, \bibinfo {author} {\bibfnamefont {E.}~\bibnamefont {Wajnryb}},
  \bibinfo {author} {\bibfnamefont {J.}~\bibnamefont {B{\l}awzdziewicz}},
  \bibinfo {author} {\bibfnamefont {J.~K.~G.}\ \bibnamefont {Dhont}}, \ and\
  \bibinfo {author} {\bibfnamefont {P.}~\bibnamefont {Lang}},\ }\href@noop {}
  {\bibfield  {journal} {\bibinfo  {journal} {J. Chem. Phys.}\ }\textbf
  {\bibinfo {volume} {132}},\ \bibinfo {pages} {074704} (\bibinfo {year}
  {2010})}\BibitemShut {NoStop}%
\bibitem [{\citenamefont {Loppinet}\ \emph {et~al.}(2012)\citenamefont
  {Loppinet}, \citenamefont {Dhont},\ and\ \citenamefont {Lang}}]{Dhont2012}%
  \BibitemOpen
  \bibfield  {author} {\bibinfo {author} {\bibfnamefont {B.}~\bibnamefont
  {Loppinet}}, \bibinfo {author} {\bibfnamefont {J.~K.~G.}\ \bibnamefont
  {Dhont}}, \ and\ \bibinfo {author} {\bibfnamefont {P.}~\bibnamefont {Lang}},\
  }\href@noop {} {\bibfield  {journal} {\bibinfo  {journal} {Eur. Phys. J. E}\
  }\textbf {\bibinfo {volume} {35}},\ \bibinfo {pages} {62} (\bibinfo {year}
  {2012})}\BibitemShut {NoStop}%
\bibitem [{\citenamefont {Michailidou}\ \emph {et~al.}(2013)\citenamefont
  {Michailidou}, \citenamefont {Swan}, \citenamefont {Brady},\ and\
  \citenamefont {Petekidis}}]{Michailidou2013}%
  \BibitemOpen
  \bibfield  {author} {\bibinfo {author} {\bibfnamefont {V.~N.}\ \bibnamefont
  {Michailidou}}, \bibinfo {author} {\bibfnamefont {J.~W.}\ \bibnamefont
  {Swan}}, \bibinfo {author} {\bibfnamefont {J.~F.}\ \bibnamefont {Brady}}, \
  and\ \bibinfo {author} {\bibfnamefont {G.}~\bibnamefont {Petekidis}},\
  }\href@noop {} {\bibfield  {journal} {\bibinfo  {journal} {J. Chem. Phys.}\
  }\textbf {\bibinfo {volume} {139}},\ \bibinfo {pages} {164905} (\bibinfo
  {year} {2013})}\BibitemShut {NoStop}%
\bibitem [{\citenamefont {Carbajal-Tinoco}\ \emph {et~al.}(1996)\citenamefont
  {Carbajal-Tinoco}, \citenamefont {Castro-Roman},\ and\ \citenamefont
  {Arauz-Lara}}]{CarbajalTinoco1996}%
  \BibitemOpen
  \bibfield  {author} {\bibinfo {author} {\bibfnamefont {M.~D.}\ \bibnamefont
  {Carbajal-Tinoco}}, \bibinfo {author} {\bibfnamefont {F.}~\bibnamefont
  {Castro-Roman}}, \ and\ \bibinfo {author} {\bibfnamefont {J.~L.}\
  \bibnamefont {Arauz-Lara}},\ }\href@noop {} {\bibfield  {journal} {\bibinfo
  {journal} {Phys. Rev. E}\ }\textbf {\bibinfo {volume} {53}},\ \bibinfo
  {pages} {3745} (\bibinfo {year} {1996})}\BibitemShut {NoStop}%
\bibitem [{\citenamefont {Schmidt}\ and\ \citenamefont
  {L\"owen}(1997)}]{Schmidt1997}%
  \BibitemOpen
  \bibfield  {author} {\bibinfo {author} {\bibfnamefont {M.}~\bibnamefont
  {Schmidt}}\ and\ \bibinfo {author} {\bibfnamefont {H.}~\bibnamefont
  {L\"owen}},\ }\href@noop {} {\bibfield  {journal} {\bibinfo  {journal} {Phys.
  Rev. E}\ }\textbf {\bibinfo {volume} {55}},\ \bibinfo {pages} {7228}
  (\bibinfo {year} {1997})}\BibitemShut {NoStop}%
\bibitem [{\citenamefont {Zangi}\ and\ \citenamefont {Rice}(2000)}]{Zangi2000}%
  \BibitemOpen
  \bibfield  {author} {\bibinfo {author} {\bibfnamefont {R.}~\bibnamefont
  {Zangi}}\ and\ \bibinfo {author} {\bibfnamefont {S.~A.}\ \bibnamefont
  {Rice}},\ }\href@noop {} {\bibfield  {journal} {\bibinfo  {journal} {Phys.
  Rev. E}\ }\textbf {\bibinfo {volume} {61}},\ \bibinfo {pages} {660} (\bibinfo
  {year} {2000})}\BibitemShut {NoStop}%
\bibitem [{\citenamefont {Frydel}\ and\ \citenamefont
  {Rice}(2003)}]{Frydel2003}%
  \BibitemOpen
  \bibfield  {author} {\bibinfo {author} {\bibfnamefont {D.}~\bibnamefont
  {Frydel}}\ and\ \bibinfo {author} {\bibfnamefont {S.~A.}\ \bibnamefont
  {Rice}},\ }\href@noop {} {\bibfield  {journal} {\bibinfo  {journal} {Phys.
  Rev. E}\ }\textbf {\bibinfo {volume} {68}},\ \bibinfo {pages} {061405}
  (\bibinfo {year} {2003})}\BibitemShut {NoStop}%
\bibitem [{\citenamefont {Han}\ \emph {et~al.}(2008)\citenamefont {Han},
  \citenamefont {Shokef}, \citenamefont {Alsayed}, \citenamefont {Yunker},
  \citenamefont {Lubensky},\ and\ \citenamefont {Yodh}}]{Han2008}%
  \BibitemOpen
  \bibfield  {author} {\bibinfo {author} {\bibfnamefont {Y.}~\bibnamefont
  {Han}}, \bibinfo {author} {\bibfnamefont {Y.}~\bibnamefont {Shokef}},
  \bibinfo {author} {\bibfnamefont {A.~M.}\ \bibnamefont {Alsayed}}, \bibinfo
  {author} {\bibfnamefont {P.}~\bibnamefont {Yunker}}, \bibinfo {author}
  {\bibfnamefont {T.~C.}\ \bibnamefont {Lubensky}}, \ and\ \bibinfo {author}
  {\bibfnamefont {A.~G.}\ \bibnamefont {Yodh}},\ }\href@noop {} {\bibfield
  {journal} {\bibinfo  {journal} {Nature}\ }\textbf {\bibinfo {volume} {456}},\
  \bibinfo {pages} {898} (\bibinfo {year} {2008})}\BibitemShut {NoStop}%
\bibitem [{\citenamefont {Lin}\ \emph {et~al.}(2000)\citenamefont {Lin},
  \citenamefont {Yu},\ and\ \citenamefont {Rice}}]{Lin2000}%
  \BibitemOpen
  \bibfield  {author} {\bibinfo {author} {\bibfnamefont {B.~H.}\ \bibnamefont
  {Lin}}, \bibinfo {author} {\bibfnamefont {J.}~\bibnamefont {Yu}}, \ and\
  \bibinfo {author} {\bibfnamefont {S.~A.}\ \bibnamefont {Rice}},\ }\href@noop
  {} {\bibfield  {journal} {\bibinfo  {journal} {Phys. Rev. E}\ }\textbf
  {\bibinfo {volume} {62}},\ \bibinfo {pages} {3909} (\bibinfo {year}
  {2000})}\BibitemShut {NoStop}%
\bibitem [{\citenamefont {Dufresne}\ \emph {et~al.}(2001)\citenamefont
  {Dufresne}, \citenamefont {Altman},\ and\ \citenamefont
  {Grier}}]{Dufresne2001}%
  \BibitemOpen
  \bibfield  {author} {\bibinfo {author} {\bibfnamefont {E.~R.}\ \bibnamefont
  {Dufresne}}, \bibinfo {author} {\bibfnamefont {D.}~\bibnamefont {Altman}}, \
  and\ \bibinfo {author} {\bibfnamefont {D.~G.}\ \bibnamefont {Grier}},\
  }\href@noop {} {\bibfield  {journal} {\bibinfo  {journal} {Europhys. Lett.}\
  }\textbf {\bibinfo {volume} {53}},\ \bibinfo {pages} {264} (\bibinfo {year}
  {2001})}\BibitemShut {NoStop}%
\bibitem [{\citenamefont {Ekiel-Je\.{z}ewska}\ \emph
  {et~al.}(2008)\citenamefont {Ekiel-Je\.{z}ewska}, \citenamefont {Wajnryb},
  \citenamefont {B{\l}awzdziewicz},\ and\ \citenamefont
  {Feuillebois}}]{Ekiel_Jezewska-Wajnryb-Blawzdziewicz-Feuillebois:2008}%
  \BibitemOpen
  \bibfield  {author} {\bibinfo {author} {\bibfnamefont {M.}~\bibnamefont
  {Ekiel-Je\.{z}ewska}}, \bibinfo {author} {\bibfnamefont {E.}~\bibnamefont
  {Wajnryb}}, \bibinfo {author} {\bibfnamefont {J.}~\bibnamefont
  {B{\l}awzdziewicz}}, \ and\ \bibinfo {author} {\bibfnamefont
  {F.}~\bibnamefont {Feuillebois}},\ }\href@noop {} {\bibfield  {journal}
  {\bibinfo  {journal} {J. Chem. Phys.}\ }\textbf {\bibinfo {volume} {129}},\
  \bibinfo {pages} {181102} (\bibinfo {year} {2008})}\BibitemShut {NoStop}%
\bibitem [{\citenamefont {Cui}\ \emph {et~al.}(2004)\citenamefont {Cui},
  \citenamefont {Diamant}, \citenamefont {Lin},\ and\ \citenamefont
  {Rice}}]{Cui2004}%
  \BibitemOpen
  \bibfield  {author} {\bibinfo {author} {\bibfnamefont {B.}~\bibnamefont
  {Cui}}, \bibinfo {author} {\bibfnamefont {H.}~\bibnamefont {Diamant}},
  \bibinfo {author} {\bibfnamefont {B.}~\bibnamefont {Lin}}, \ and\ \bibinfo
  {author} {\bibfnamefont {S.~A.}\ \bibnamefont {Rice}},\ }\href@noop {}
  {\bibfield  {journal} {\bibinfo  {journal} {Phys. Rev. Lett.}\ }\textbf
  {\bibinfo {volume} {92}},\ \bibinfo {pages} {258301} (\bibinfo {year}
  {2004})}\BibitemShut {NoStop}%
\bibitem [{\citenamefont {Bhattacharya}\ \emph
  {et~al.}(2005{\natexlab{a}})\citenamefont {Bhattacharya}, \citenamefont
  {B\l{}awzdziewicz},\ and\ \citenamefont {Wajnryb}}]{Bhattacharya2005b}%
  \BibitemOpen
  \bibfield  {author} {\bibinfo {author} {\bibfnamefont {S.}~\bibnamefont
  {Bhattacharya}}, \bibinfo {author} {\bibfnamefont {J.}~\bibnamefont
  {B\l{}awzdziewicz}}, \ and\ \bibinfo {author} {\bibfnamefont
  {E.}~\bibnamefont {Wajnryb}},\ }\href@noop {} {\bibfield  {journal} {\bibinfo
   {journal} {J. Fluid Mech.}\ }\textbf {\bibinfo {volume} {541}},\ \bibinfo
  {pages} {263} (\bibinfo {year} {2005}{\natexlab{a}})}\BibitemShut {NoStop}%
\bibitem [{\citenamefont {Diamant}\ \emph {et~al.}(2005)\citenamefont
  {Diamant}, \citenamefont {Cui}, \citenamefont {Lin},\ and\ \citenamefont
  {Rice}}]{Diamant2005A}%
  \BibitemOpen
  \bibfield  {author} {\bibinfo {author} {\bibfnamefont {H.}~\bibnamefont
  {Diamant}}, \bibinfo {author} {\bibfnamefont {B.}~\bibnamefont {Cui}},
  \bibinfo {author} {\bibfnamefont {B.}~\bibnamefont {Lin}}, \ and\ \bibinfo
  {author} {\bibfnamefont {S.~A.}\ \bibnamefont {Rice}},\ }\href@noop {}
  {\bibfield  {journal} {\bibinfo  {journal} {J. Phys.: Condens. Matter}\
  }\textbf {\bibinfo {volume} {17}},\ \bibinfo {pages} {S4047} (\bibinfo {year}
  {2005})}\BibitemShut {NoStop}%
\bibitem [{\citenamefont {B{\l}awzdziewicz}\ and\ \citenamefont
  {Wajnryb}(2012)}]{Blawzdziewicz-Wajnryb:2012}%
  \BibitemOpen
  \bibfield  {author} {\bibinfo {author} {\bibfnamefont {J.}~\bibnamefont
  {B{\l}awzdziewicz}}\ and\ \bibinfo {author} {\bibfnamefont {E.}~\bibnamefont
  {Wajnryb}},\ }\href@noop {} {\bibfield  {journal} {\bibinfo  {journal} {J.
  Phys. Conf. Ser.}\ }\textbf {\bibinfo {volume} {392}},\ \bibinfo {pages}
  {012008} (\bibinfo {year} {2012})}\BibitemShut {NoStop}%
\bibitem [{\citenamefont {Baron}\ \emph {et~al.}(2008)\citenamefont {Baron},
  \citenamefont {B{\l}awzdziewicz},\ and\ \citenamefont
  {Wajnryb}}]{Baron-Blawzdziewicz-Wajnryb:2008}%
  \BibitemOpen
  \bibfield  {author} {\bibinfo {author} {\bibfnamefont {M.}~\bibnamefont
  {Baron}}, \bibinfo {author} {\bibfnamefont {J.}~\bibnamefont
  {B{\l}awzdziewicz}}, \ and\ \bibinfo {author} {\bibfnamefont
  {E.}~\bibnamefont {Wajnryb}},\ }\href@noop {} {\bibfield  {journal} {\bibinfo
   {journal} {Phys. Rev. Lett.}\ }\textbf {\bibinfo {volume} {100}},\ \bibinfo
  {pages} {174502} (\bibinfo {year} {2008})}\BibitemShut {NoStop}%
\bibitem [{\citenamefont {Lin}\ \emph {et~al.}(1995)\citenamefont {Lin},
  \citenamefont {Rice},\ and\ \citenamefont {Weitz}}]{Lin1995}%
  \BibitemOpen
  \bibfield  {author} {\bibinfo {author} {\bibfnamefont {B.}~\bibnamefont
  {Lin}}, \bibinfo {author} {\bibfnamefont {S.~A.}\ \bibnamefont {Rice}}, \
  and\ \bibinfo {author} {\bibfnamefont {D.~A.}\ \bibnamefont {Weitz}},\
  }\href@noop {} {\bibfield  {journal} {\bibinfo  {journal} {Phys. Rev. E}\
  }\textbf {\bibinfo {volume} {51}},\ \bibinfo {pages} {423} (\bibinfo {year}
  {1995})}\BibitemShut {NoStop}%
\bibitem [{\citenamefont {Cichocki}\ \emph {et~al.}(2004)\citenamefont
  {Cichocki}, \citenamefont {Ekiel-Je\.{z}ewska}, \citenamefont {Nagele},\ and\
  \citenamefont {Wajnryb}}]{Cichocki2004}%
  \BibitemOpen
  \bibfield  {author} {\bibinfo {author} {\bibfnamefont {B.}~\bibnamefont
  {Cichocki}}, \bibinfo {author} {\bibfnamefont {M.~L.}\ \bibnamefont
  {Ekiel-Je\.{z}ewska}}, \bibinfo {author} {\bibfnamefont {G.}~\bibnamefont
  {Nagele}}, \ and\ \bibinfo {author} {\bibfnamefont {E.}~\bibnamefont
  {Wajnryb}},\ }\href@noop {} {\bibfield  {journal} {\bibinfo  {journal} {J.
  Chem. Phys.}\ }\textbf {\bibinfo {volume} {121}},\ \bibinfo {pages} {2305}
  (\bibinfo {year} {2004})}\BibitemShut {NoStop}%
\bibitem [{\citenamefont {Peng}\ \emph {et~al.}(2009)\citenamefont {Peng},
  \citenamefont {Chen}, \citenamefont {Fischer}, \citenamefont {Weitz},\ and\
  \citenamefont {Tong}}]{Peng2009}%
  \BibitemOpen
  \bibfield  {author} {\bibinfo {author} {\bibfnamefont {Y.}~\bibnamefont
  {Peng}}, \bibinfo {author} {\bibfnamefont {W.}~\bibnamefont {Chen}}, \bibinfo
  {author} {\bibfnamefont {T.~M.}\ \bibnamefont {Fischer}}, \bibinfo {author}
  {\bibfnamefont {D.~A.}\ \bibnamefont {Weitz}}, \ and\ \bibinfo {author}
  {\bibfnamefont {P.}~\bibnamefont {Tong}},\ }\href@noop {} {\bibfield
  {journal} {\bibinfo  {journal} {J. Fluid Mech.}\ }\textbf {\bibinfo {volume}
  {618}},\ \bibinfo {pages} {243} (\bibinfo {year} {2009})}\BibitemShut
  {NoStop}%
\bibitem [{\citenamefont {Zhang}\ \emph {et~al.}(2014)\citenamefont {Zhang},
  \citenamefont {Chen}, \citenamefont {Li}, \citenamefont {Zhang},\ and\
  \citenamefont {Chen}}]{Zhang2014}%
  \BibitemOpen
  \bibfield  {author} {\bibinfo {author} {\bibfnamefont {W.}~\bibnamefont
  {Zhang}}, \bibinfo {author} {\bibfnamefont {S.}~\bibnamefont {Chen}},
  \bibinfo {author} {\bibfnamefont {N.}~\bibnamefont {Li}}, \bibinfo {author}
  {\bibfnamefont {J.~Z.}\ \bibnamefont {Zhang}}, \ and\ \bibinfo {author}
  {\bibfnamefont {W.}~\bibnamefont {Chen}},\ }\href@noop {} {\bibfield
  {journal} {\bibinfo  {journal} {{PLoS ONE}}\ }\textbf {\bibinfo {volume}
  {{9}}} (\bibinfo {year} {{2014}})}\BibitemShut {NoStop}%
\bibitem [{\citenamefont {Skinner}\ \emph {et~al.}(2010)\citenamefont
  {Skinner}, \citenamefont {Aarts},\ and\ \citenamefont
  {Dullens}}]{Skinner2010}%
  \BibitemOpen
  \bibfield  {author} {\bibinfo {author} {\bibfnamefont {T.~O.~E.}\
  \bibnamefont {Skinner}}, \bibinfo {author} {\bibfnamefont {D.~G. A.~L.}\
  \bibnamefont {Aarts}}, \ and\ \bibinfo {author} {\bibfnamefont {R.~P.~A.}\
  \bibnamefont {Dullens}},\ }\href@noop {} {\bibfield  {journal} {\bibinfo
  {journal} {Phys. Rev. Lett.}\ }\textbf {\bibinfo {volume} {105}},\ \bibinfo
  {pages} {168301} (\bibinfo {year} {2010})}\BibitemShut {NoStop}%
\bibitem [{\citenamefont {Kapfenberger}\ \emph {et~al.}(2013)\citenamefont
  {Kapfenberger}, \citenamefont {Sonn-Segev},\ and\ \citenamefont
  {Roichman}}]{Kapfenberger2013}%
  \BibitemOpen
  \bibfield  {author} {\bibinfo {author} {\bibfnamefont {D.}~\bibnamefont
  {Kapfenberger}}, \bibinfo {author} {\bibfnamefont {A.}~\bibnamefont
  {Sonn-Segev}}, \ and\ \bibinfo {author} {\bibfnamefont {Y.}~\bibnamefont
  {Roichman}},\ }\href@noop {} {\bibfield  {journal} {\bibinfo  {journal} {Opt.
  Express}\ }\textbf {\bibinfo {volume} {21}},\ \bibinfo {pages} {12228}
  (\bibinfo {year} {2013})}\BibitemShut {NoStop}%
\bibitem [{\citenamefont {Lee}\ and\ \citenamefont {Grier}(2007)}]{Lee07a}%
  \BibitemOpen
  \bibfield  {author} {\bibinfo {author} {\bibfnamefont {S.-H.}\ \bibnamefont
  {Lee}}\ and\ \bibinfo {author} {\bibfnamefont {D.~G.}\ \bibnamefont
  {Grier}},\ }\href@noop {} {\bibfield  {journal} {\bibinfo  {journal} {Opt.
  Express}\ }\textbf {\bibinfo {volume} {15}},\ \bibinfo {pages} {1505}
  (\bibinfo {year} {2007})}\BibitemShut {NoStop}%
\bibitem [{\citenamefont {Cheong}\ \emph {et~al.}(2010)\citenamefont {Cheong},
  \citenamefont {Krishnatreya},\ and\ \citenamefont {Grier}}]{Cheong2010b}%
  \BibitemOpen
  \bibfield  {author} {\bibinfo {author} {\bibfnamefont {F.~C.}\ \bibnamefont
  {Cheong}}, \bibinfo {author} {\bibfnamefont {B.~J.}\ \bibnamefont
  {Krishnatreya}}, \ and\ \bibinfo {author} {\bibfnamefont {D.~G.}\
  \bibnamefont {Grier}},\ }\href@noop {} {\bibfield  {journal} {\bibinfo
  {journal} {Optics Express}\ }\textbf {\bibinfo {volume} {18}},\ \bibinfo
  {pages} {13563} (\bibinfo {year} {2010})}\BibitemShut {NoStop}%
\bibitem [{\citenamefont {Behrens}\ and\ \citenamefont
  {Grier}(2001)}]{Behrens2001}%
  \BibitemOpen
  \bibfield  {author} {\bibinfo {author} {\bibfnamefont {S.~H.}\ \bibnamefont
  {Behrens}}\ and\ \bibinfo {author} {\bibfnamefont {D.~G.}\ \bibnamefont
  {Grier}},\ }\href@noop {} {\bibfield  {journal} {\bibinfo  {journal} {The
  Journal of Chemical Physics}\ }\textbf {\bibinfo {volume} {115}},\ \bibinfo
  {pages} {6716} (\bibinfo {year} {2001})}\BibitemShut {NoStop}%
\bibitem [{\citenamefont {Israelachvili}(1992)}]{Israelachvili}%
  \BibitemOpen
  \bibfield  {author} {\bibinfo {author} {\bibfnamefont {J.~N.}\ \bibnamefont
  {Israelachvili}},\ }\href@noop {} {\emph {\bibinfo {title} {Intermolecular
  and Surface Forces}}}\ (\bibinfo  {publisher} {Academic Press, London},\
  \bibinfo {year} {1992})\BibitemShut {NoStop}%
\bibitem [{\citenamefont {Frenkel}\ and\ \citenamefont
  {Smit}(2002)}]{Frenkel-Smit:2002}%
  \BibitemOpen
  \bibfield  {author} {\bibinfo {author} {\bibfnamefont {D.}~\bibnamefont
  {Frenkel}}\ and\ \bibinfo {author} {\bibfnamefont {B.}~\bibnamefont {Smit}},\
  }\href@noop {} {\emph {\bibinfo {title} {{Understanding Molecular Simulation.
  From Algorithms to Simulations}}}}\ (\bibinfo  {publisher} {Academic Press},\
  \bibinfo {address} {New York},\ \bibinfo {year} {2002})\BibitemShut {NoStop}%
\bibitem [{\citenamefont {B{\l}awzdziewicz}\ and\ \citenamefont
  {Wajnryb}(2008)}]{Blawzdziewicz-Wajnryb:2008}%
  \BibitemOpen
  \bibfield  {author} {\bibinfo {author} {\bibfnamefont {J.}~\bibnamefont
  {B{\l}awzdziewicz}}\ and\ \bibinfo {author} {\bibfnamefont {E.}~\bibnamefont
  {Wajnryb}},\ }\href@noop {} {\bibfield  {journal} {\bibinfo  {journal} {Phys.
  Fluids.}\ }\textbf {\bibinfo {volume} {20}},\ \bibinfo {pages} {093303}
  (\bibinfo {year} {2008})}\BibitemShut {NoStop}%
\bibitem [{\citenamefont {Bhattacharya}\ \emph
  {et~al.}(2005{\natexlab{b}})\citenamefont {Bhattacharya}, \citenamefont
  {B{\l}awzdziewicz},\ and\ \citenamefont
  {Wajnryb}}]{Bhattacharya-Blawzdziewicz-Wajnryb:2005a}%
  \BibitemOpen
  \bibfield  {author} {\bibinfo {author} {\bibfnamefont {S.}~\bibnamefont
  {Bhattacharya}}, \bibinfo {author} {\bibfnamefont {J.}~\bibnamefont
  {B{\l}awzdziewicz}}, \ and\ \bibinfo {author} {\bibfnamefont
  {E.}~\bibnamefont {Wajnryb}},\ }\href@noop {} {\bibfield  {journal} {\bibinfo
   {journal} {Physica A}\ }\textbf {\bibinfo {volume} {356}},\ \bibinfo {pages}
  {294} (\bibinfo {year} {2005}{\natexlab{b}})}\BibitemShut {NoStop}%
\bibitem [{\citenamefont {Bhattacharya}\ \emph
  {et~al.}(2005{\natexlab{c}})\citenamefont {Bhattacharya}, \citenamefont
  {B{\l}awzdziewicz},\ and\ \citenamefont
  {Wajnryb}}]{Bhattacharya-Blawzdziewicz-Wajnryb:2005}%
  \BibitemOpen
  \bibfield  {author} {\bibinfo {author} {\bibfnamefont {S.}~\bibnamefont
  {Bhattacharya}}, \bibinfo {author} {\bibfnamefont {J.}~\bibnamefont
  {B{\l}awzdziewicz}}, \ and\ \bibinfo {author} {\bibfnamefont
  {E.}~\bibnamefont {Wajnryb}},\ }\href@noop {} {\bibfield  {journal} {\bibinfo
   {journal} {J. Fluid Mech.}\ }\textbf {\bibinfo {volume} {541}},\ \bibinfo
  {pages} {263} (\bibinfo {year} {2005}{\natexlab{c}})}\BibitemShut {NoStop}%
\bibitem [{\citenamefont {Cichocki}\ and\ \citenamefont
  {Felderhof}(1989)}]{Cichocki-Felderhof:1989}%
  \BibitemOpen
  \bibfield  {author} {\bibinfo {author} {\bibfnamefont {B.}~\bibnamefont
  {Cichocki}}\ and\ \bibinfo {author} {\bibfnamefont {B.~U.}\ \bibnamefont
  {Felderhof}},\ }\href@noop {} {\bibfield  {journal} {\bibinfo  {journal}
  {Physica A}\ }\textbf {\bibinfo {volume} {159}},\ \bibinfo {pages} {19}
  (\bibinfo {year} {1989})}\BibitemShut {NoStop}%
\bibitem [{\citenamefont {Sadlej}\ \emph {et~al.}(2009)\citenamefont {Sadlej},
  \citenamefont {Wajnryb}, \citenamefont {B{\l}awzdziewicz}, \citenamefont
  {Ekiel-Je\.{z}ewska},\ and\ \citenamefont
  {Adamczyk}}]{Sadlej-Wajnryb-Blawzdziewicz-Ekiel_Jezewska-Adamczyk:2009}%
  \BibitemOpen
  \bibfield  {author} {\bibinfo {author} {\bibfnamefont {K.}~\bibnamefont
  {Sadlej}}, \bibinfo {author} {\bibfnamefont {E.}~\bibnamefont {Wajnryb}},
  \bibinfo {author} {\bibfnamefont {J.}~\bibnamefont {B{\l}awzdziewicz}},
  \bibinfo {author} {\bibfnamefont {M.}~\bibnamefont {Ekiel-Je\.{z}ewska}}, \
  and\ \bibinfo {author} {\bibfnamefont {Z.}~\bibnamefont {Adamczyk}},\
  }\href@noop {} {\bibfield  {journal} {\bibinfo  {journal} {J. Chem. Phys.}\
  }\textbf {\bibinfo {volume} {130}},\ \bibinfo {pages} {144706} (\bibinfo
  {year} {2009})}\BibitemShut {NoStop}%
\bibitem [{\citenamefont {Abade}\ \emph {et~al.}(2010)\citenamefont {Abade},
  \citenamefont {Cichocki}, \citenamefont {Ekiel-Je\.{z}ewska}, \citenamefont
  {Naegele},\ and\ \citenamefont {Wajnryb}}]{Abade:2010}%
  \BibitemOpen
  \bibfield  {author} {\bibinfo {author} {\bibfnamefont {G.}~\bibnamefont
  {Abade}}, \bibinfo {author} {\bibfnamefont {B.}~\bibnamefont {Cichocki}},
  \bibinfo {author} {\bibfnamefont {M.~L.}\ \bibnamefont {Ekiel-Je\.{z}ewska}},
  \bibinfo {author} {\bibfnamefont {G.}~\bibnamefont {Naegele}}, \ and\
  \bibinfo {author} {\bibfnamefont {E.}~\bibnamefont {Wajnryb}},\ }\href@noop
  {} {\bibfield  {journal} {\bibinfo  {journal} {J. Chem. Phys.}\ }\textbf
  {\bibinfo {volume} {132}},\ \bibinfo {pages} {014503} (\bibinfo {year}
  {2010})}\BibitemShut {NoStop}%
\bibitem [{\citenamefont {Crocker}\ and\ \citenamefont
  {Grier}(1996)}]{Crocker1996}%
  \BibitemOpen
  \bibfield  {author} {\bibinfo {author} {\bibfnamefont {J.~C.}\ \bibnamefont
  {Crocker}}\ and\ \bibinfo {author} {\bibfnamefont {D.~G.}\ \bibnamefont
  {Grier}},\ }\href@noop {} {\bibfield  {journal} {\bibinfo  {journal} {J.
  Colloid. Interf. Sci.}\ }\textbf {\bibinfo {volume} {179}},\ \bibinfo {pages}
  {298} (\bibinfo {year} {1996})}\BibitemShut {NoStop}%
\bibitem [{\citenamefont {Helfand}\ \emph {et~al.}(1961)\citenamefont
  {Helfand}, \citenamefont {Frisch},\ and\ \citenamefont
  {Lebowitz}}]{Helfand-Frisch-Lebowitz:1961}%
  \BibitemOpen
  \bibfield  {author} {\bibinfo {author} {\bibfnamefont {E.}~\bibnamefont
  {Helfand}}, \bibinfo {author} {\bibfnamefont {H.~L.}\ \bibnamefont {Frisch}},
  \ and\ \bibinfo {author} {\bibfnamefont {J.~L.}\ \bibnamefont {Lebowitz}},\
  }\href@noop {} {\bibfield  {journal} {\bibinfo  {journal} {J. Chem. Phys.}\
  }\textbf {\bibinfo {volume} {34}},\ \bibinfo {pages} {1037} (\bibinfo {year}
  {1961})}\BibitemShut {NoStop}%
\bibitem [{\citenamefont {Chen}\ and\ \citenamefont {Ma}(2006)}]{Chen2006}%
  \BibitemOpen
  \bibfield  {author} {\bibinfo {author} {\bibfnamefont {H.}~\bibnamefont
  {Chen}}\ and\ \bibinfo {author} {\bibfnamefont {H.}~\bibnamefont {Ma}},\
  }\href@noop {} {\bibfield  {journal} {\bibinfo  {journal} {J. Chem. Phys.}\
  }\textbf {\bibinfo {volume} {125}},\ \bibinfo {pages} {024510} (\bibinfo
  {year} {2006})}\BibitemShut {NoStop}%
\end{thebibliography}
\end{document}